\documentclass{article}%
\usepackage[T1]{fontenc}%
\usepackage{float}%
\usepackage{geometry}
\usepackage{graphicx}%
\usepackage[colorlinks=true,linkcolor=blue,citecolor=blue]{hyperref}%
\usepackage{caption}%
\usepackage{natbib}%
\usepackage{amsmath}%
\usepackage{amssymb}%
\usepackage{mathtools}%
\usepackage{booktabs}%
\usepackage{caption}%
\usepackage{subcaption}%
\usepackage{framed}%
\usepackage{rotating}
\usepackage{adjustbox}

\usepackage{makecell}

\newcommand{\EE}{\mathbb{E}}
\newcommand{\hpi}{\widehat{\pi}}
\newcommand{\htree}{\hat{\pi}^{\text{C}}}
\newcommand{\hfixed}{\hat{\pi}^{\text{N}}}

\title{Contextual Bandits in a Survey Experiment on Charitable Giving: Within-Experiment Outcomes versus Policy Learning}
\author{Susan Athey, Undral Byambadalai, Vitor Hadad, \\ Sanath Kumar Krishnamurthy, Weiwen Leung, Joseph Jay Williams \thanks{
Authors are listed in alphabetical order. We are grateful for the generous financial support provided by the Sloan Foundation, the Golub Capital Social Impact Lab at Stanford Graduate School of Business, and the Office of Naval Research grant N00014-19-1-2468. Molly Offer-Westort and Imanol Ibarra provided excellent advice and research assistance on this or earlier versions of the project. Athey: Stanford Graduate School of Business. Byambadalai: Stanford Graduate School of Business. Hadad: Amazon. Krishnamurthy: Stanford Management Science and Engineering. Weiwen Leung: Coupang. Joseph Jay Williams: University of Toronto Computer Science.}}
\date{\today}

\begin{document}

\maketitle

\begin{abstract}
We design and implement an adaptive experiment (a ``contextual bandit'') to learn a targeted treatment assignment policy, where the goal is to use a participant's survey responses to determine which charity to expose them to in a donation solicitation. The design balances two competing objectives: optimizing the outcomes for the subjects in the experiment (``cumulative regret minimization'') and gathering data that will be most useful for policy learning, that is, for learning an assignment rule that will maximize welfare if used after the experiment (``simple regret minimization''). We evaluate alternative experimental designs by collecting pilot data and then conducting a simulation study. Next, we implement our selected algorithm. Finally, we perform a second simulation study anchored to the collected data that evaluates the benefits of the algorithm we chose. Our first result is that the value of a learned policy in this setting is higher when data is collected via a uniform randomization rather than collected adaptively using standard cumulative regret minimization or policy learning algorithms. We propose a simple heuristic for adaptive experimentation that improves upon uniform randomization from the perspective of policy learning at the expense of increasing cumulative regret relative to alternative bandit algorithms. The heuristic modifies an existing contextual bandit algorithm by (i) imposing a lower bound on assignment probabilities that decay slowly so that no arm is discarded too quickly, and (ii) after adaptively collecting data, restricting policy learning to select from arms where sufficient data has been gathered.  
\end{abstract}

\section{Introduction}

Randomized experiments that contrast several treatments to one another (and possibly to a control group) can be used not just to understand the average differences between treatment arms but also to inform the design of targeted treatment assignment policies - a task we refer to as ``policy learning.'' For example, if the problem concerns which drug to give to a patient, policy learning entails estimating a rule that assigns the best drug to patients given their characteristics. When an experiment has more than a few arms, however, it can be expensive or time consuming to gather enough data to accurately estimate the performance of each treatment arm for subgroups of the population. In this setting, adaptive experiments can be used to focus data collection on promising treatment arms for each set of participant characteristics. Thus, the researcher may learn a more effective targeted treatment assignment policy (or learn a good policy more quickly). Experimental subjects simultaneously benefit as they are more likely to be exposed to treatments that work for them during the experiment. Assigning treatments in this way reduces ethical concerns related to assigning participants to clearly inferior treatment arms. 

However, adaptive experiments also have several potential downsides. For example, treatment arms might be discarded too quickly. Further, it may be difficult at the end of the experiment to accurately compare alternative (counterfactual) targeted treatment assignment policies if some arms are assigned with low probability in some regions of the participant characteristic space. Indeed, adaptive experimentation where treatment assignment varies with participant characteristics creates just the types of problems for estimation (assignment probabilities that vary with characteristics and may be very low) that a large literature on estimating treatment effects with observational data attempts to address \citep{imbens2015causal}. Because learning a good policy requires evaluating a variety of alternative policies that may not have much overlap with the policies used to collect data (``off-policy evaluation''), adaptive data collection can increase the chance of mistakes in policy learning \citep{kitagawa2018should, athey2021policy, zhou2018offline}. Adaptive data collection can also raise challenges for hypothesis testing after the experiment, as discussed in, e.g., \cite{hadad2021confidence}, \cite{zhan2021off}, and \cite{zhan2021policy}.

In this paper, we consider the problem of designing an adaptive experiment with two goals. The first goal is to use the data from the experiment to design an effective targeted treatment assignment rule at the end of the experiment.  Our second goal is to avoid assigning observations to suboptimal treatment arms during the experiment, and in particular, to target treatment assignments to individuals on the basis of their observed characteristics (their ``contexts''). This type of adaptive experiment is known as a ``contextual bandit.'' In our study, this implies minimizing assignment to proposed charities for participants identified to be unlikely to respond well to them. This goal is often important in settings where the treatments make a big difference to participants. This consideration is substantively less important in our application. Another reason this goal can be important is because there is an opportunity cost in making a solicitation due to limited attention as well as the potential opportunity cost of the space on a web page where a donation is solicited. An objective in this paper is to illustrate the trade-off between the goal of maximizing outcomes for the participants in the experiment and our first goal of policy learning.

We apply our proposed model in an experiment where the treatment is the choice of which charity to show to a participant, and the outcome is how participants felt about making a donation to the selected charity (a hypothetical proxy for how much they would donate). In our experiment, we gathered information about participant characteristics from a set of survey questions. Then, we used those characteristics to determine which treatment, or charity, to show the participant. We find that when targeting is not possible, Greenpeace is the optimal charity to recommend to individuals. When targeting is possible, the policy selects Greenpeace for many participants and selects more polarizing charities for specific subgroups. However, even when targeting is possible, it is difficult to find a subgroup and a charity that strongly dominates Greenpeace - other than the group who feels strongly about the right to bear arms for whom the National Rifle Association (NRA) is more appealing. Targeting based on identification of participants being liberal versus conservative has relatively modest benefits. Assigning Planned Parenthood instead of Greenpeace to liberals increases their average outcome by 7.2\%, and assigning NRA instead of Greenpeace to conservatives increases their average outcome by 2.6\% (See Table \ref{tab:subgroup_policy_values}). We used the data we collected to create semi-synthetic data sets for simulation studies.  We used these data sets to establish that, in our setting, the approach of running a contextual bandit algorithm is expected to lead to a slightly (about 1\%) better learned policy than collecting data uniformly using a standard randomized controlled trial, while at the same time substantially improving the experiment's outcomes. 

Our paper is one of a small number (e.g., \cite{li2010contextual} and \cite{bastani2021efficient}) that has conducted a contextual bandit in practice and analyzed its performance relative to alternatives (Section \ref{sec:RelLit} reviews the literature).  The relative lack of empirical applications of this type contrasts with the massive applied literature that estimates causal effects of treatments using observational data. When comparing estimation methods, a single data set can be used to compare a variety of methods. In contrast, contextual bandit algorithms are algorithms that guide data collection. They require a particular arm to be used for a particular covariate vector (context), so that it is not straightforward to reanalyze historically collected data to simulate the performance of a bandit. Given that running survey experiments on popular platforms such as Mechanical Turk or Lucid can cost $\$1$ or more per subject, it is difficult to run many parallel experiments comparing algorithms with sufficient sample size.  Thus, papers comparing methods have typically used some form of simulation or semi-synthetic data. 

We address the challenge of comparing contextual bandit methods and evaluating the trade-offs in their performance for different objectives by combining several rounds of real-world data collection with semi-synthetic simulations.  We proceed in three steps: First, we run a pilot and use it to create semi-synthetic data for comparing and selecting among algorithms; Second, we run a single contextual bandit algorithm on real subjects using the selected algorithm; Third, we conduct a second simulation exercise using the adaptively-collected data to update the parameters used to generate semi-synthetic data. 

To describe the problem of adaptive experimental design more precisely, some basic notation is helpful. Each observation $t$ is represented by $(X_t, W_t, Y_t)$, where $X_t \in \mathbb{R}^{p}$ is the \emph{context}, a vector of observed covariates (demographic characteristics, measures of political affiliation, etc.); $W_t \in [K]$ \footnote{$[K]$ denotes the set $\{1,2, \cdots, K\}$.} is the index of the treatment arm assigned to that participant; and $Y_t$ is the observed outcome. We let $Y_t(1), Y_t(2), \cdots, Y_t(K)$ denote potential outcomes under treatment arm $\{1,2,\cdots, K\}$, so that $Y_t = Y_t(W_t)$. 

The first goal described above is \emph{policy learning} \citep[see e.g.,][]{manski2004statistical, stoye2009minimax, kitagawa2018should, athey2021policy}. A \emph{policy} is a deterministic treatment rule, formalized as a function from contexts to treatments $\pi: \mathbb{R}^{p} \mapsto [K]$. In the policy learning problem, we are given some previously collected data (following any arbitrary assignment rule) and a set of available policies $\Pi$. Our objective is to approximate the policy $\pi^*$ that maximizes the expected outcome, i.e., $$\pi^* := \arg\max_{\pi \in \Pi} \EE[Y_t(\pi(X_t))].$$
The set of available policies is often restricted in some way, such as belonging to the set of decision-tree policies \citep[e.g.,][]{zhou2018offline} or satisfying some cost-benefit threshold \citep[e.g.,][]{sun2021empirical}. We can also express the goal as selecting a policy $\hat\pi$ based on available data that minimizes expected \emph{simple regret}, or the expected loss from using the learned policy in the future:
$$R(\hat\pi)=\EE[Y_t(\pi^*(X_t))]-\EE[Y_t(\hat\pi(X_t))].$$

The second goal is often called \emph{cumulative regret minimization}. ``Cumulative regret'' refers to the gap in expected reward attained by the treatment $W_t$ that was assigned to a participant (e.g., in an adaptive experiment) and the one that would have been attained under the optimal policy $\pi^*$, i.e., $\EE[Y_t(\pi^*(X_t))] - \EE[Y_t(W_t)]$.\footnote{Some authors call this \emph{expected regret} and reserve the term regret to mean the random variable $Y_t(\pi^*(X_t)) - Y_t(W_t)$.} Of course, minimizing cumulative regret also implies maximizing rewards accrued during the experiment. There is a large literature in adaptive experimental design that deals with the cumulative regret minimization problem \citep[see e.g.,][Chapters 18-32]{lattimore2020bandit}. Such algorithms, referred to as contextual bandit algorithms, devise optimal strategies to minimize the time spent acting somewhat at random (``exploration'') to quickly learn which treatments work best and focus assignment on those treatments (``exploitation''). Often these strategies are probabilistic: Given a vector of covariates $x_t$, the algorithm outputs a vector of probabilities from which the arm identity is drawn.

At first glance, the two aforementioned goals might seem aligned. To learn which policies work, the researcher needs ``enough'' data about the performance of each arm across the covariate space (i.e., for different types of participants), but, at the same time, it seems wasteful to collect ``too much'' data on arms that are clearly suboptimal. Since suboptimal arms are disfavored by a cumulative regret minimization algorithm, one could hypothesize that such an algorithm would also generate data that would be amenable for policy learning. And indeed, these forces can lead algorithms focused on cumulative regret to also improve on nonadaptive experiments in terms of policy learning (\cite{even2006action,kasy2021adaptive}).

However, it turns out that the two goals are partially in conflict. As we demonstrate using simulations designed based on our pilot data, the true value of the best policies estimated from adaptive data collection is often lower than that of the best policies estimated from nonadaptive data collection, such as through a randomized control trial with uniform assignment probabilities. We hypothesize that the problem is that standard bandit algorithms are overly aggressive, dropping arms too quickly for a policy learning method to do a good job at approximating the optimal policy. 

The above issue motivates us to propose an algorithm based on simple heuristics that align both goals - achieving high quality policy learning while also minimizing regret during the experiment. We begin with a cumulative regret minimization algorithm but modify it in two ways to achieve both of our goals. First, we impose a lower bound on assignment probabilities suggested by the algorithm. This lower bound decays slowly, so that no arm is discarded too quickly. Second, when learning a policy at the end of the adaptive experiment, we compute a score that tracks how much each arm is favored by the cumulative regret minimization algorithm. We then learn a policy using only the arms that are most favored. We show that this heuristic is able to learn a policy whose value on average is at least as good as the one we would have learned by collecting data non-adaptively (i.e., via a randomized controlled trial with uniform assignment probabilities) while still yielding substantial reduction in regret accrued during the experiment.

%Incidentally, one may also ask why not just keep the assignment rule that was being used by the cumulative regret minimization algorithm at the end of the adaptive data collection. There are two reasons. First, the policy is often not deterministic, since as mentioned above cumulative regret minimization algorithms often employ randomized strategies, and it's not clear how exactly to collapse the randomness into a well-defined function. Second, even if we found a way to make the randomized rule deterministic, we cannot guarantee that the resulting policy would belong to the desired policy class $\Pi$.

We compare five distinct contextual bandit algorithms in our simulation study to evaluate the trade-offs between cumulative and simple regret (i.e., in-experiment outcomes versus policy learning) in these commonly-used algorithms. The first is \textit{Uniform}, where all treatments are assigned with equal probability, as in a classic randomized controlled trial. Next, we consider three different bandit algorithms proposed in the literature. A classic algorithm is \textit{Linear Thompson sampling}, where at each stage, a regression is used to fit an outcome model as a function of the treatment arm and participant characteristics. This model in turn is used to construct, conditional on a vector of participant characteristics, the posterior probability that arm \textit{w} is best. Arms are assigned to participants according to the posterior probability that the arm is optimal.  We also consider two modifications of this algorithm that have been shown theoretically to optimize for policy learning: Exploration Sampling \citep{kasy2021adaptive} and Top-Two Thompson Sampling \citep{russo2016simple}. We show that in the empirical setting we study, the theoretical promise of these algorithms is not born out in practice; they do not in fact improve over Uniform assignment for policy learning. To address the problems we find with these algorithms, we propose a final algorithm we call \textit{TreeBagging}, which is inspired by bagging algorithms studied by, e.g., \cite{agarwal2014taming}. With \textit{TreeBagging}, assignment probabilities are determined by aggregating the results of a procedure that repeatedly estimates tree-based treatment assignment policies learned on training data sets drawn with replacement from previous observations. We incorporate the heuristic modification discussed above (a decaying lower bound on assignment probabilities, combined with selection of arms for policy learning) to \textit{TreeBagging}.

The rest of the paper is structured as follows: Section \ref{sec:RelLit} reviews related literature. Section \ref{sec:research} describes the research design. Section \ref{sec:design} describes the main experiment and the pilot experiments conducted to select the bandit algorithm most appropriate for our setting and objective, including the Treebagging algorithm we propose in detail. Section \ref{sec:parameter} demonstrates the performance of the Treebagging algorithm through simulations and compares adaptive and nonadaptive experiments. Section \ref{sec:survey} describes our survey design, while Section \ref{sec:implementation} explains how our experiments were implemented. Section \ref{sec:main} shows the results of a new experiment conducted on recruited subjects in which we deployed the Treebagging algorithm. Section \ref{sec:FinalSimul} compares bandit performance with that of uniform randomization. Section \ref{sec:conclusion} concludes. Additional figures and tables, as well as mathematical details, are provided in the Appendix.

\section{Related Literature} \label{sec:RelLit}
Our work falls within a small but growing literature on adaptive experimental design that considers the goal of policy learning (equivalently, ``minimizing simple regret''). Several algorithms have been proposed and analyzed theoretically. For example, for the problem without contexts, \cite{kasy2021adaptive} propose a modification of Thompson sampling called ``exploration sampling,'' which assigns treatments in proportion to the posterior probability that an arm is optimal; this modification slows down learning relative to Thompson sampling and performs well at policy learning. \cite{deshmukh2018simple} propose a ``contextual gap'' algorithm, where the assignment probability for a given arm and context is inversely proportional to the gap between the estimated best possible payoff for the context and the estimated payoff for a given arm. Theoretical performance and simulations indicate the algorithm shows promise at policy learning. \cite{chambaz2017targeted} consider a setting in which there are covariates but only two arms and propose a sequential experiment design for nonparametric policy learning; their method is based on assigning observations to treatment or control depending on how certain one is of a positive treatment effect. The authors show that under specific exploration regimes, they can learn an optimal policy with sufficient data.  The TreeBagging algorithm we propose in this paper attempts to explicitly balance the two objectives of policy learning and cumulative regret.

The \emph{best arm identification} literature studies a problem related to the one we address in this paper. This literature aims to identify the best treatment arm in a setting with no covariates as quickly as possible or with the highest possible certainty  \citep[see e.g.,][]{audibert2010best, jamieson2014best, russo2016simple, kasy2021adaptive,lu2021variance}. Relaxing the requirement of finding the best arm to an approximate requirement, the $(\epsilon,\delta)$-PAC best arm identification problem that considers the problem of identifying, with probability at least $1-\delta$, an arm that is within $\epsilon$ of being optimal, with as few samples as possible \citep[see e.g.,][]{even2006action,hassidim2020optimal}. Another related line of work studies best arm identification in a multi-armed bandit setting that places additional structure on the problem. This setting assumes that the expected reward of arms (as a function of arm features) lies in a known function class. The best arm identification in this setting exploits this additional structure with bounds that depend on the complexity of the reward model class \citep[see e.g.,][]{soare2014best,xu2018fully,jedra2020optimal,kazerouni2021best}.

The practical limitations of algorithms designed for policy learning have not been fully explored in the literature. In particular, we are not aware of research documenting findings analogous to our result that in a range of outcome models consistent with our pilot data, uniform (nonadaptive) sampling performs better than algorithms that were explicitly designed for policy learning, though the fact that uniform sampling is in general a strong baseline is suggested by the theoretical results in \cite{bubeck2009pure}. The findings in this paper suggest that there is still room for improvement in algorithms designed for policy learning, and also for algorithms that optimize for regret in a contextual bandit settings; see \cite{krishnamurthy2021adapting, krishnamurthy2021optimal, carranza2022flexible, simchi2022bypassing} for some recent discussions of challenges for standard algorithms that arise when the functional form of the mapping between individual characteristics and outcomes is unknown.  

Although a full theoretical analysis is beyond the scope of this paper, here we show that a few modifications of standard algorithms improve performance. In particular, we incorporate lower bounds on assignment probabilities in our proposed TreeBagging algorithm. The idea of modifying a cumulative regret minimization algorithm by imposing a lower bound on assignment probabilities to attain better policy learning is motivated by proposals such as the Tempered Thompson algorithm described in \cite{caria2020adaptive}.  

Our paper also contributes to a small but growing literature using contextual bandits in real-world applications.
Bandit algorithms have been used in a wide range of applications including mobile health interventions \citep{rabbi2019optimizing}, clinical trial design \citep{durand2018contextual}, news article recommendation \citep{li2010contextual}, and public health interventions \citep{mate2021field}. However, there are only a handful of studies that apply contextual bandits to real-world participants (rather than in simulations) and test hypotheses about the impact of the final policy.  \cite{bastani2021efficient} show the usefulness of adaptive design based on real-time data in informing border testing policies during the COVID-19 pandemic in Greece by comparing the value of the targeted policy with those of counterfactual policies. \cite{yang2020targeting} study optimal targeted discounts for The Boston Globe subscribers and implemented their learned targeted policy with bootstrap Thompson sampling in a second experiment.  \cite{offer2021optimal} study effective interventions to combat the spread of misinformation about COVID-19 on social media in sub-Saharan Africa by learning and evaluating a targeted policy. \cite{caria2020adaptive} propose the Tempered Thompson algorithm, which balances the goals of maximizing the precision of treatment effects estimates and maximizing the welfare of experiment participants; they implemented the method to study labor market policies in Jordan.

From a substantive perspective, our paper estimates a targeted treatment assignment policy for targeting specific charities on the basis of participant characteristics. We are not aware of other papers addressing this type of objective in charitable giving. However, the targeting policies we learn are not particularly surprising, and indeed the charities (e.g., the NRA) were selected to ensure that there would be a strong benefit to targeting. Nonetheless, the magnitudes of the treatment effects we estimate and how they differ by subgroup may be useful to charitable giving platforms and organizations. The broad problem of encouraging giving has been studied from different angles in economics and other social sciences. See \cite{andreoni2006philanthropy}, \cite{list2011market} and \cite{andreoni2013charitable} for a summary on general facts about charitable giving and a review of theoretical and empirical research within public economics.

\section{Research Design}\label{sec:research}

\begin{table}[H]
    \centering
    \begin{tabular}{lccc}
    \hline
    \hline
        Step & Goal & \thead{Data collection method/ \\ data used} & Algorithm(s) \\
        \hline
        Pilot 1 &  & \thead{Uniform randomization\\(nonadaptive)} &  \thead{None} \\
        Pilot 2 &  & \thead{Adaptive} & \thead{TreeBagging(50)} \\
        \makecell[l]{Simulation study 1: \\ Bandit tuning} & \thead{Tune bandit parameters \\ and compare algorithms} & \thead{Semi-synthetic data \\ based on Pilot 2 data} & \thead{Uniform\\ TreeBagging(50)\\ BootstrapThompson\\ BootstrapES\\ BootstrapTTTS}\\
        \makecell[l]{Main experiment: \\ Learning phase} & \thead{Learn best contextual\\ \& non-contextual policies} &  \thead{Adaptive} & \thead{TreeBagging(50)} \\
        \makecell[l]{Main experiment:\\ Evaluation phase}  & \thead{Evaluate performance of\\ best contextual \&\\ non-contextual policies\\ from learning phase} &  \thead{Nonadaptive \\ and nonuniform} & \thead{None} \\
        \makecell[l]{Simulation study 2: \\ Update semi-synthetic data} & \thead{Update parameters used to \\ generate semi-synthetic data \\ and compare algorithms} & \thead{Semi-synthetic data \\ based on Pilot 2\\ \& main experiment data} & \thead{Uniform\\ TreeBagging(50)\\ BootstrapThompson\\ BootstrapES\\ BootstrapTTTS}\\
        \hline
        \hline
    \end{tabular}
    \caption{Research design}
    \label{tab:research}
\end{table}

Table \ref{tab:research} provides an overview of the overall research design, including two pilots, a simulation study used to select tuning parameters for the adaptive experiment, the two phases of the main experiment, and a final simulation study conducted using the data from the main experiment which revisits questions about the benefits of adaptivity.

\subsection{Survey Design} \label{sec:survey}

For the Pilots and the Main Experiment, we implemented survey experiments using recruited subjects (Mechanical Turk for the pilots, and Lucid for the Main Experiment). In each case, the survey experiment consisted of three sections. In the first section, participants answered questions about general demographic characteristics, political affiliation, and media consumption. In the second section, participants were shown a charity logo and a paragraph of information about it. They were asked to rank, on a scale from -10 to 10, ``\textit{how would people like you feel if we donated one thousand dollars to the following charity}.''\footnote{This framing of the question was designed to measure participants' preferences for charities, given budget constraints and the need to avoid deception.} The charity shown to the participant was selected uniformly at random from a list of eight charities: National Rifle Association (NRA), the American Israel Public Affairs Committee (AIPAC), Greenpeace, Planned Parenthood Federation of America, the Black Lives Matter movement (BLM), the Chan Zuckerberg Initiative (CZI), People for the Ethical Treatment of Animals (PETA), and the Clinton Foundation. The display contained the charity's logo and a description (from the organization's Wikipedia page). 

The questions regarding participant characteristics were chosen to capture features hypothesized to influence preferences for different charities. The end goal was an experiment where targeted treatment assignment would result in higher average outcomes than a policy where all individuals were exposed to the same charity. 

\subsection{Pilot Experiments and Implementation}
\label{sec:implementation}

Before the main experiment, we conducted two pilot experiments to finalize our experimental design. The first pilot experiment, which was conducted in March of 2020, was a randomized experiment in which the participants were recruited via Amazon Mechanical Turk and were randomly assigned to one of the treatments with equal probability. We call this experiment the Pilot 1 experiment (n=2,463). The second pilot experiment, which was conducted in August of 2021, was an adaptive experiment with a learning and evaluation phase. We call this experiment the Pilot 2 experiment (n=3,064). In Pilot 2, the participants were recruited via Lucid. More survey details and the analyses of the pilot experiments can be found in Appendices \ref{sec:contexts} and \ref{sec:pilot}.

In Pilots 1 and 2, we included the Salvation Army as one of the charities, bringing the total number of treatments (charities) to nine. Pilot 1 data revealed that the Salvation Army is favored by both conservatives and liberals (see Tables \ref{tab:pilot_charity_per_leaning} and \ref{tab:pilot_charity_per_leaning_lucid} in Appendix \ref{sec:pilot}), and it was selected as the best fixed policy in Pilot 2 (see Table \ref{tab:pilot_learned_policies_lucid} in Appendix \ref{sec:pilot}). The contextual policies of maximal depth 1 to 3 that are learned in Pilot 2 are not statistically distinguishable from assigning everyone to the Salvation Army (also in Table \ref{tab:pilot_learned_policies_lucid} in Appendix \ref{sec:pilot}). As different groups showed strong preferences for the Salvation Army, we excluded the Salvation Army as a treatment in the main experiment in order to illustrate gains from personalizing treatments when there is an underlying heterogeneity in optimal treatment.

\subsection{Main Experiment}
\label{sec:design}
Our main experiment has two phases: the learning phase and the evaluation phase. The goal of the \textit{learning} phase is to learn a treatment assignment policy, balancing exploration and exploitation with both cumulative regret and policy learning. Participants are recruited in batches during this phase, with the treatment assignment probabilities in each batch changing based on the results of the previous batch. In this experiment, the assignment probabilities for each batch are determined using the TreeBagging algorithm we detail in the next subsection. 

At the end of the learning phase, we use the collected data from all batches to estimate two different types of policies. The first is the best non-contextual policy $\hfixed$; that is, the best single treatment arm if we had to assign the same treatment arm to all participants. The second estimated policy is a contextual (tree) policy for treatment assignment $\htree$, where subgroups can be assigned different treatments based on a decision \textit{tree} that forks according to a participant's covariates. 

The primary goal of the \textit{evaluation} phase is to provide accurate estimates of the policies learned at the end of the learning phase. In particular, we examine whether the estimated contextual (tree) policy (personalized treatment assignment) is better than the estimated non-contextual policy (assignment to the best single treatment).

During the evaluation phase, we collect data non-adaptively (but also not uniformly) using the following procedure. With probability $\epsilon$, which we set to $\epsilon=0.3$, treatment arms are uniformly randomly assigned; with probability $(1-\epsilon)/2$, we assign arms according to the learned tree policy $\htree$; and with probability $(1-\epsilon)/2$, we assign arms according to the learned non-contextual policy $\hfixed$. The uniform sampling with probability $\epsilon$ is to gather data for additional analyses after the experiment.

\subsubsection{TreeBagging Algorithm for the Learning Phase} 

As we discuss further below, simulations based on pilot data showed poor performance for standard learning algorithms, and so we designed a new algorithm, referred to as the TreeBagging Algorithm, that performed better.\footnote{See, for example, \cite{krishnamurthy2021adapting, krishnamurthy2021optimal, carranza2022flexible, simchi2022bypassing} for recent discussions of the challenges with selecting functional forms in contextual bandits, as well as proposals for algorithms that may perform better.} In the learning phase, we compute an assignment probability function for each batch according to a standard \emph{bagging} algorithm \citep[see e.g.,][]{agarwal2014taming} to which we make two changes. We first describe a bagging algorithm generally, then detail the changes. 

In essence, a bagging algorithm, or bootstrap aggregating,  entails drawing samples of observations with replacement. Specifically, at the beginning of a new batch, we estimate a sequence of policies $\hat{\pi}^{(s)}(x)$ for $s \in [S]$ batches (we set $S=50$), where $x_t \in \mathbb{R}^{p}$ are participant covariates. Each policy in $S$ is fit by sampling with replacement from previous batches' data.\footnote{The first batch uses uniform random assignment.} Then, for each value of $x_t$ observed in the new batch, we compute tentative assignment probabilities for each arm $w_t$ according to the proportion of fitted policies in the ensemble of policies to which observations with a value of $x$ were assigned. Letting $\tilde{e}_t(x_t, w_t)$ denote the assignment probabilities suggested by the algorithm for participant $t$, 
\begin{equation}
    \label{eq:bagging_probs}
    \tilde{e}_t(x_t, w_t) \propto | \{ \hat{\pi}^{(s)}(x_t) = w_t  \}|.
\end{equation}

In terms of obtaining high rewards \emph{during} the experiment (i.e., cumulative regret minimization), the bagging algorithm described above often performs similarly to other optimal bandit algorithms \citep{bietti2018contextual}. However, as we demonstrate in the next section, when we use data collected via this heuristic to learn a policy (i.e., simple regret minimization), we often end up with a policy whose value is relatively low (i.e., the approach is not successful at identifying the best policy assignment rule, which is one of our stated goals). In fact, we are often better off collecting data by assigning arms uniformly at random. 

We propose two adjustments to this bagging algorithm. First, our simulations suggest that we can often obtain a higher-value policy and reduce the likelihood of successful treatment arms being discarded by imposing a decaying lower bound $f(t) := t^{-\alpha}/K \: (\alpha > 0, K > 0)$ on the assignment probabilities above:
\begin{align}
    \label{eq:probs_with_floor}
    e_t(x, w) := 
    \begin{cases}
    f(t) \qquad &\text{if } \tilde{e}_t(x, w) < f(t), \\
      f(t)  + c(\tilde{e}_t(x, w) - f(t)) \qquad &\text{otherwise},
    \end{cases}
\end{align}
where $c > 0$ is a number chosen so that all probabilities sum to one. When a new batch arrives, the values of $ f(t), \tilde{e}_t(x, w)$ and $c$ in \eqref{eq:probs_with_floor} are recomputed. 

The second adjustment occurs at the end of the learning phase, when we use the data collected thus far to learn a contextual tree policy. To promote overlap between the data collected and the learned policy, we drop arms that are not promising. We choose which arms to drop by assigning the following ``frequency score'' that shows how often a treatment arm is assigned:
\begin{equation}
    \label{eq:freq}
    \text{freq}(w) := 
    \frac{1}{S} \sum_{s=1}^S \frac{1}{T_{\text{learn}}} 
    \sum_{t=1}^{T_{\text{learn}}} \mathbb{I}
        \left\{\hpi_B^{(s)}(X_t) = w \right\},
\end{equation}
where $T_{\text{learn}}$ is the length of the learning batch, and $\pi_B^{(s)}$ are the policies in the bagging ensemble during the last learning phase batch. We order arms according to their score \eqref{eq:freq} in descending order and select only the top $k$ arms. Our default value is $k=4$, and our experiment has eight arms (charities), so half of all arms are selected. 

Once we have selected the top arms, we use the subset of data assigned to these arms to compute Augmented Inverse Propensity Weighting scores (AIPW; see Appendix \ref{sec:aipw}) for each observation and estimate a tree policy, following \cite{athey2021policy}. The tree depth (i.e., the highest number of nodes from the root of the tree to the bottommost tree leaf) is chosen via cross-validation. We fit tree policies of depths 1 and 2 on the first 80\% of the data (training data) and then evaluate these policies on the remaining 20\% of the data (test data). We select the tree depth that yields the highest estimated value in the test data. Finally,  we refit a policy with the selected depth using the entire data set (training plus test data). This final policy is denoted by $\htree$. Figure \ref{fig:learned_policy} in Section \ref{sec:main} illustrates a learned contextual policy tree. 

%In addition to the contextual policy described above, we also learn a non-contextual policy that assigns a single arm, regardless of the context. This policy is denoted by $\hfixed$.

%\paragraph{Evaluation phase} The goal of the evaluation phase is to provide accurate estimates of the policies learned at the end of the learning phase. Therefore, our data-collection procedure is described as follows. An $\epsilon$ fraction of the time we assign arms uniformly at random; the remaining $1 - \epsilon$ half is split equally into assignment according to the learned tree policy $\htree$ and assignment according to the learned non-contextual policy $\hfixed$. The parameter $\epsilon$ is tuned via simulations in the next section.

\subsection{Bandit Tuning}
\label{sec:parameter}

This section outlines the simulations undertaken to determine our chosen algorithm and the parameter values in Table \ref{tab:parameters} that were used in the main experiment. Our choices aim to balance several objectives: high per-period value obtained during the experiment (low cumulative regret), high policy value for the policy estimated at the end of the experiment (policy learning), and accurate estimates of the benefit of adaptive experimentation.

\begin{table}[H]
    \centering
    \begin{tabular}{l|l}  \toprule
         Parameter & Selected value \\ \midrule
         Total length of experiment &  3,000 \\
         Fraction of data collected during the \emph{learning} phase &  1/2 (1,500 periods) \\
         Decay rate on assignment probabilities lower bound ($\alpha$) & 1/16 \\
         Number of arms selected via frequency score for Treebagging ($k$) & 4  \\ \bottomrule
    \end{tabular}
    \caption{Algorithm parameters and their default values}
    \label{tab:parameters}
\end{table}

\paragraph{Simulation design} We use the data from Pilot 2 for the simulations, which we refer to as ``pilot data'' throughout this section for simplicity. To generate contexts, we draw with replacement from pilot data. To generate outcomes, we proceed as follows. First, we draw a sample with replacement from the pilot data. 
Next, we fit an ordinal classification model to predict the probability that a participant will choose each response level, given contexts (e.g., age, gender and political leaning) and treatments (charities).\footnote{The model used here is based on the penalized logistic regression generalization for ordinal outcomes described in \cite{rennie2005loss} and implemented in the Python package \texttt{mord}.} 
The set of explanatory variables included covariates (and their squares), treatments, and interactions between covariates and dummified treatments. The regularization parameter used when fitting the model above (i.e., the factor multiplying the sum of squares of coefficients in the regularized regression) impacts the heterogeneity in the data and is chosen randomly from a set of prespecified parameters $\{10,50,100,500\}$ ranging from low to high.\footnote{Higher regularization parameters induce lower heterogeneity in our data-generating process. We consider fairly high regularization parameters to allow for data-generating processes that are relatively pessimistic in how much heterogeneity is induced. } 

%These values are chosen so that the average learned contextual policy attains a value that is between 0.8 and 1.1 points above the value non-contextual policy at the default parameter configuration explained below. Recalling from Table \ref{tab:pilot_learned_policies} that, according to our pilot data, contextual policies attain at least 1.1 points more than the the non-contextual policy, this implies that our data-generating process is relatively pessimistic in how much heterogeneity it induces.

\paragraph{Algorithms} We consider different treatment assignment schemes. \textit{(i)} \emph{Uniform}, in which all treatments are assigned with equal probability, i.e., $e_t(x, w) \equiv 1/K$; \textit{(ii)} \emph{TreeBagging}, following the explanation in the previous section; \textit{(iii)} \emph{Linear Thompson sampling}, in which we maintain an approximate posterior probability that each arm is optimal conditional on contexts, based on a linear model of the outcome, and assign treatments roughly according to this posterior. That is, if according to our model, the Bayesian posterior probability that arm $w$ is the best is $x\%$ for observation $t$, then we assign that arm $w$ with $x\%$ probability; \textit{(iv)} \emph{Exploration Sampling}, an Exploration Sampling variant of Algorithm (iii); \textit{(v)} \emph{Top-Two Thompson Sampling}, a Top-Two Thompson Sampling variant of Algorithm (iii). Algorithms (ii) and (iii) are formally described in Appendix \ref{sec:algorithms}. Table \ref{tab:parameters} describes default values of tuning parameters associated with these algorithms.

Tables \ref{tab:simulation_learned_policy_value} and \ref{tab:simulation_regret} compare simulation results across algorithms. For \emph{TreeBagging}, Table \ref{tab:simulation_learned_policy_value} shows that the learned policy is at least as good as the one learned when collecting data through uniform randomization, and, in fact, it shows improvement by a modest margin. Table \ref{tab:simulation_regret} shows that, indeed, regret decreases (by around 20\%) with \emph{TreeBagging} as compared to \emph{Uniform}. 
%Tables \ref{tab:simulation_learned_policy_value} and \ref{tab:simulation_regret} quantify the benefits of running an adaptive experiment. Tables \ref{tab:simulation_learned_policy_value} shows that the learned policy is at least as good as the one learned when collecting data non-adaptive (i.e,. uniformly), and in fact shows improves upon in by a modest margin. Table \ref{tab:simulation_regret} shows that we indeed get modest benefits (about 20\% reduction) in terms of regret. Finally, Figures \ref{fig:tradeoffs} show the trade-offs between the true value of the learned contextual policy, the power to detect a difference in value between the learned contextual and non-contextual policies, and the average regret incurred during the learning phase.

\begin{table}[H]
    \centering
    %\begin{tabular}{llrrrr}
%\toprule
%        & Heterogeneity &  High &  Medium High &  Medium Low &   Low  \\
%Bandit & {} &         &          &         &          \\
% \midrule
% Uniform & Mean &   6.027 &  5.968 &    5.908 &     5.656 \\
%         & Std. Error &   0.003 &  0.003 &  0.002 &       0.002 \\
%         &\\
% BootstrapThompson & Mean &   5.999 &  5.934 &   5.881 &      5.629 \\
%         & Std. Error &   0.003 &  0.003 &  0.002 &       0.002 \\
%         &\\
% TreeBagging(50) & Mean &   6.094 &   6.029 &  5.963 &      5.693 \\
%         & Std. Error &   0.003 &    0.002 &   0.002 &    0.002 \\

% \bottomrule
% \end{tabular}

\begin{tabular}{lcccc}
\toprule
Regularization &  10 & 50 & 100 & 500 \\
Bandit &        &          &         &          \\
\midrule
Uniform &   6.031 &    5.962 &  5.912  &    5.653 \\
        &    (0.004) &    (0.004) &   (0.004) &  (0.004) \\
&\\
TreeBagging(50) &   6.096 &    6.026 &   5.963 &    5.692 \\
        &    (0.004) &    (0.004) &   (0.004) &    (0.003) \\  
        &\\
BootstrapThompson &  5.994 &    5.938 &  5.883 &    5.623 \\
        &   (0.004) &    (0.004) &   (0.004) &    (0.004) \\
&\\
BootstrapES &   6.028 &   5.957  &  5.897 &    5.641 \\
        &   (0.004) &    (0.004) &   (0.004) &   (0.004) \\
&\\
BootstrapTTTS &    6.014 &    5.955 &  5.896 &    5.638 \\
        &   (0.004) &  (0.004) &  (0.004) &  (0.004) \\
&\\ 
Improvement  & 101.08\% & 101.07\% & 100.86\% & 100.69\% \\
(TreeBagging(50) as \% of Uniform) \\
\bottomrule
\end{tabular}
    \caption{This table shows the value of learned tree policy $\htree$ when data is collected using bandit algorithms (indicated in rows) using the default parameters in Table \ref{tab:parameters}. Each column corresponds to a regularization parameter used in fitting the outcome model that underlies our simulation of participant outcomes. Lower regularization parameters correspond to higher heterogeneity in the simulated data. Averages and standard errors (in parentheses) are computed across over 1,000 simulations per cell. The last row indicates the improvement of TreeBagging(50) over Uniform in terms of value of policy learned.}
    \label{tab:simulation_learned_policy_value}
\end{table}

\begin{table}[H]
    \centering
    % \begin{tabular}{llrrrr}
% \toprule
%       & Heterogeneity &  High &  Medium High &  Medium Low &   Low \\
% Bandit & {} &       &         &         &       \\
% \midrule
% Uniform & Mean & 5.560 & 5.249 &  5.063 &    4.451 \\
%         & Std. Error & 0.002 &   0.002 &   0.002 & 0.002 \\
%         &\\
% BootstrapThompson & Mean & 4.182 &  3.951 &  3.815 &   3.374 \\
%         & Std. Error & 0.001 &   0.001 &   0.001 & 0.001 \\
%         &\\
% TreeBagging(50) & Mean & 4.442 & 4.164 &  4.001 &    3.495 \\
%         & Std. Error & 0.002 &   0.001 &   0.001 & 0.001 \\
% \bottomrule
% \end{tabular}

\begin{tabular}{lcccc}
\toprule
Regularization &  10 & 50 & 100 & 500 \\
Bandit &  &       &         &              \\
\midrule
Uniform &  5.553 &    5.250 &   5.062 & 4.450 \\
        &  (0.003) &  (0.003) &  (0.003) & (0.002) \\
&\\
TreeBagging(50) & 4.439 &    4.164 &   4.000 & 3.494 \\
        &  (0.002) &   (0.002) &  (0.002) & (0.002) \\
&\\

BootstrapThompson & 4.179 &   3.953 &   3.814 & 3.372 \\
        &  (0.002) &   (0.002) &  (0.002) &  (0.002) \\
&\\
BootstrapES & 4.325 &    4.099 &   3.962 & 3.518 \\
        & (0.002) &  (0.002) & (0.002) & (0.002) \\
&\\
BootstrapTTTS & 4.315 &   4.092 &   3.954 & 3.509 \\
        & (0.002) & (0.002) & (0.002) & (0.002) \\
&\\        
Reduction  & 79.94\% & 79.31\% & 79.02\% & 78.52\% \\
(TreeBagging(50) as \% of Uniform)  \\
\bottomrule
\end{tabular}

    \caption{This table presents the averaged per-period regret attained while using default parameters in Table \ref{tab:parameters} where applicable under different algorithms. Each column indicates the regularization parameter used in fitting the outcome model that underlies our simulation of participant outcomes. Lower regularization parameters correspond to higher heterogeneity in the simulated data. Averages and standard errors (in parentheses)  are computed across over 1,000 simulations per cell. The last row indicates the reduction in average regret of TreeBagging(50) over Uniform.}
    \label{tab:simulation_regret}
\end{table}

\begin{figure}[H]
    \centering
    \includegraphics[width=1\textwidth]{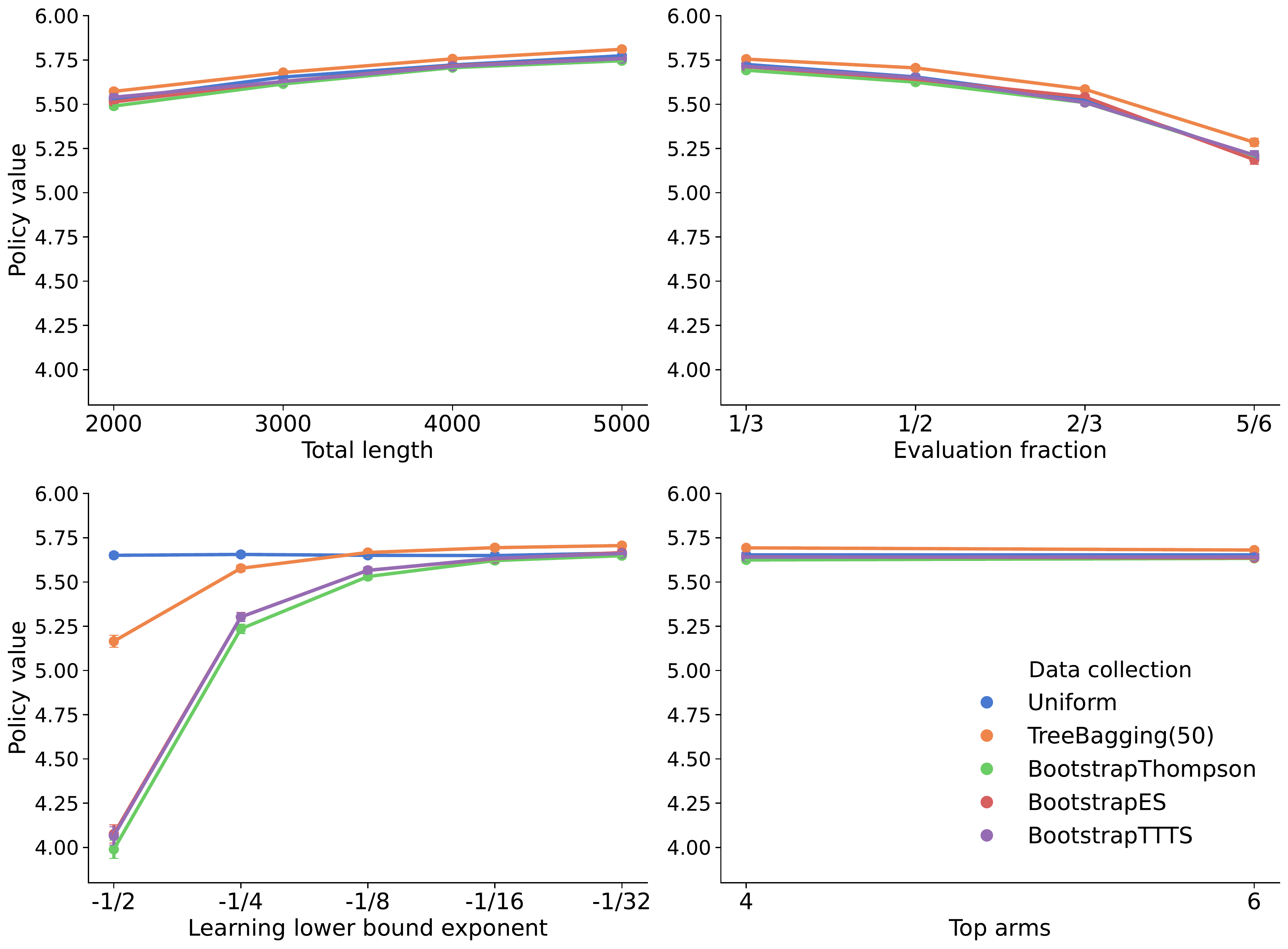}
    \caption{Each subplot shows the average value of learned policies across simulations as we vary one tuning parameter while fixing others at their default values. Error bars are 95\% confidence intervals computed across over 1,000 simulations per parameter configuration.}
    \label{fig:values_variation}
\end{figure}

Figure \ref{fig:values_variation} shows the value of the learned policy as we vary any one of the default parameters shown in Table \ref{tab:parameters}. Of particular interest is that the value of the policy increases as we increase the lower bound on assignment probabilities. This reinforces our intuition that the ``raw'' assignment probabilities output by traditional bandit algorithms are too aggressive (i.e., discards treatment arms too quickly) for accurate off-policy learning. 

\begin{figure}[H]
    \centering
    \includegraphics[width=0.24\textwidth]{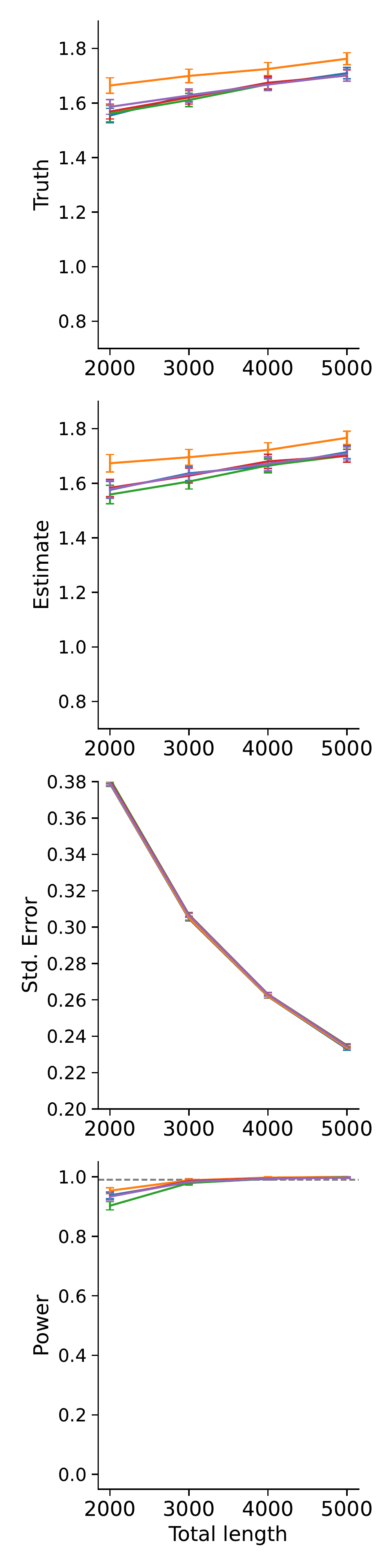}
    \includegraphics[width=0.24\textwidth]{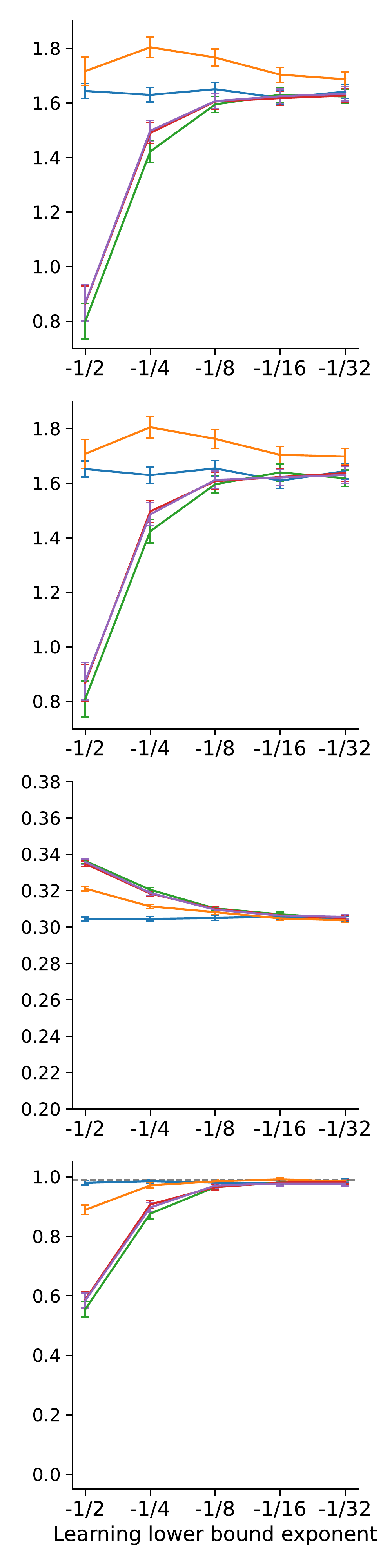}
    \includegraphics[width=0.24\textwidth]{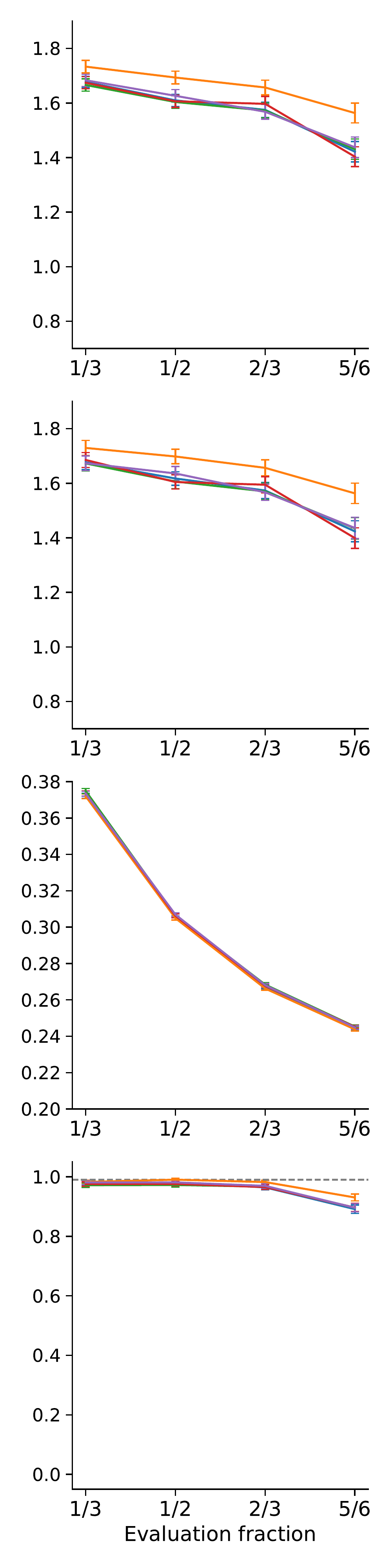}
    \includegraphics[width=0.24\textwidth]{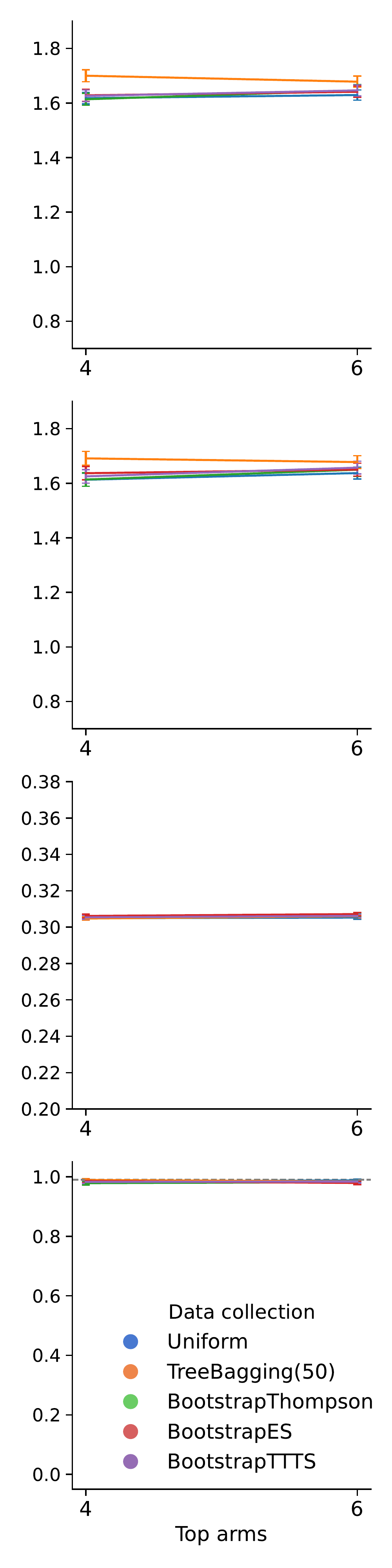}
    \caption{Each row shows statistics associated with the estimand $\EE[Y_t(\htree(X_t))] - \EE[Y_t(\hfixed(X_t))]$ (i.e., the value difference between the learned contextual and learned non-contextual policy). In each column, we vary one tuning parameter and fix others at their default values. Errors bars are 95\% confidence intervals computed across over 1,000 simulations per parameter configuration.}
    \label{fig:contrast_variation}
\end{figure}

In Figure \ref{fig:contrast_variation}, we present various statistics associated with the difference in value between the learned contextual and non-contextual policies, as we vary parameters away from their ``default'' value. We present the average of the following statistics over simulation runs: the true difference between the two learned policies, the estimate of the difference, the standard error, and the statistical power. At the default experiment length of $3,000$ observations, we have over 99\% statistical power to detect differences between these two policies. 

\begin{figure}[H]
    \centering
    \includegraphics[width=\textwidth]{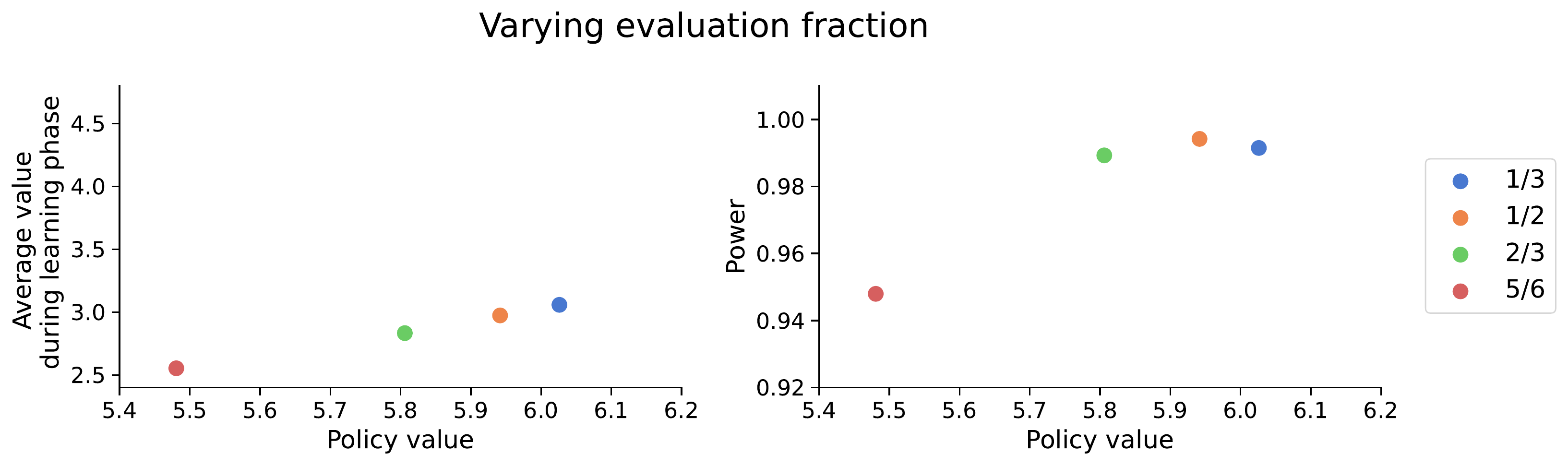}
    \includegraphics[width=\textwidth]{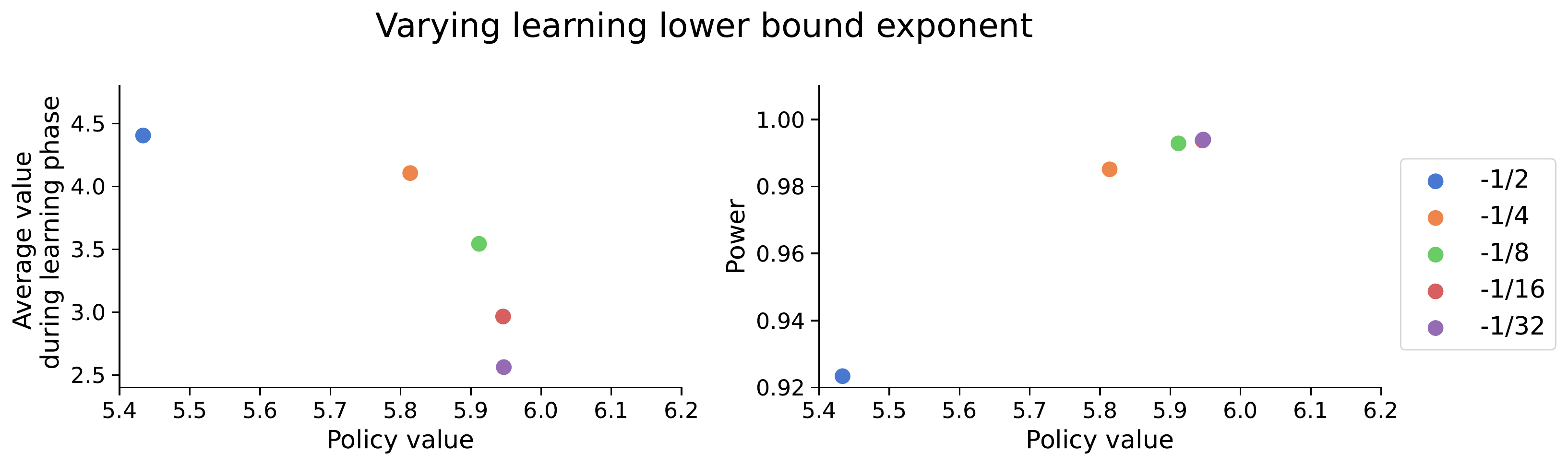}
    \includegraphics[width=\textwidth]{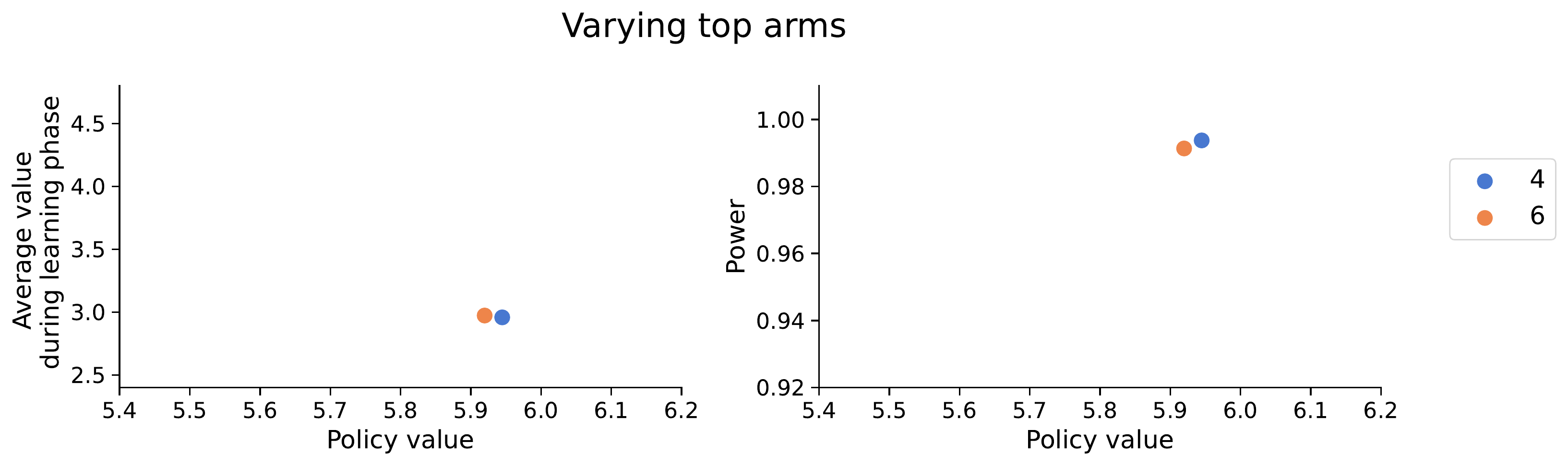}
    \includegraphics[width=\textwidth]{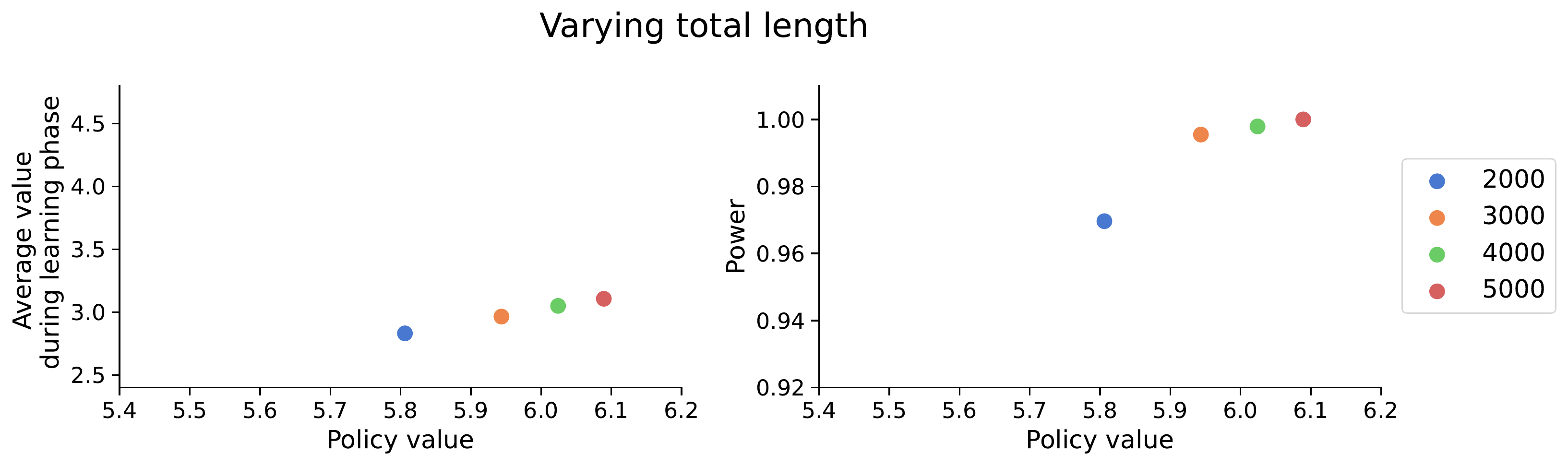}
    \caption{This figure shows trade-offs between learned contextual policy value $\EE[Y_t(\htree(X_t))]$ vs average value during learning phase (left) and learned contextual policy value $\EE[Y_t(\htree(X_t))]$ vs power associated with the hypothesis $H_{0}: \EE[Y_t(\htree(X_t))] = \EE[Y_t(\hfixed(X_t))]$ (i.e., the value difference between the learned contextual and learned non-contextual policy) (right). In each panel, we vary one tuning parameter and fix others at their default values. Estimates are computed across over 1,000 simulations per parameter configuration.}
    \label{fig:tradeoffs}
\end{figure}

Lastly, Figure \ref{fig:tradeoffs} shows the trade-offs between the true value of the learned contextual policy, the average value during the learning phase, and the power to detect a difference in value between the learned contextual and non-contextual policies. Consider first the evaluation fraction, where a value of 1/2 attains high policy value and high average value during the learning phase. Although a value of 1/3 attains both higher policy value and higher average value during the learning phase compared to 1/2, the power to detect the value difference between the learned contextual and non-contextual policies is lower. Next, consider the lower bound exponent, where a value of -1/16 attains a high policy value while attaining higher average value during the learning phase than -1/32. Selecting four arms attains higher policy value while attaining a slightly lower average value during the learning phase compared to selecting six arms, though the differences are very small. Lastly, the power to detect the difference is 97\% at length 2,000 and 99\% at length 3,000.

\section{Main Experiment Results}
\label{sec:main}

In December of 2021, we conducted our main survey experiment in which participants were recruited via Lucid. 

\paragraph{Behavior during the experiment.} Table \ref{tab:batch_n} shows the number of valid observations in each batch. We aimed to collect 150 observations in each batch, but the numbers are not exact as we do not have fine control over the flow of participants' responses in Lucid.

\begin{table}[H]
    \centering
    \begin{tabular}{lrrrrrrrrrr}
\toprule
Batch &   1  &   2  &   3  &   4  &   5  &   6  &   7  &   8  &   9  &   10 \\
\midrule
Number of obs. &  192 &  145 &  141 &  145 &  174 &  149 &  147 &  165 &  149 &  153 \\
\bottomrule
\end{tabular}

    \caption{Number of valid observations in each batch}
    \label{tab:batch_n}
\end{table}

\begin{figure}[H]
    \centering
    \includegraphics[width=.45\textwidth]{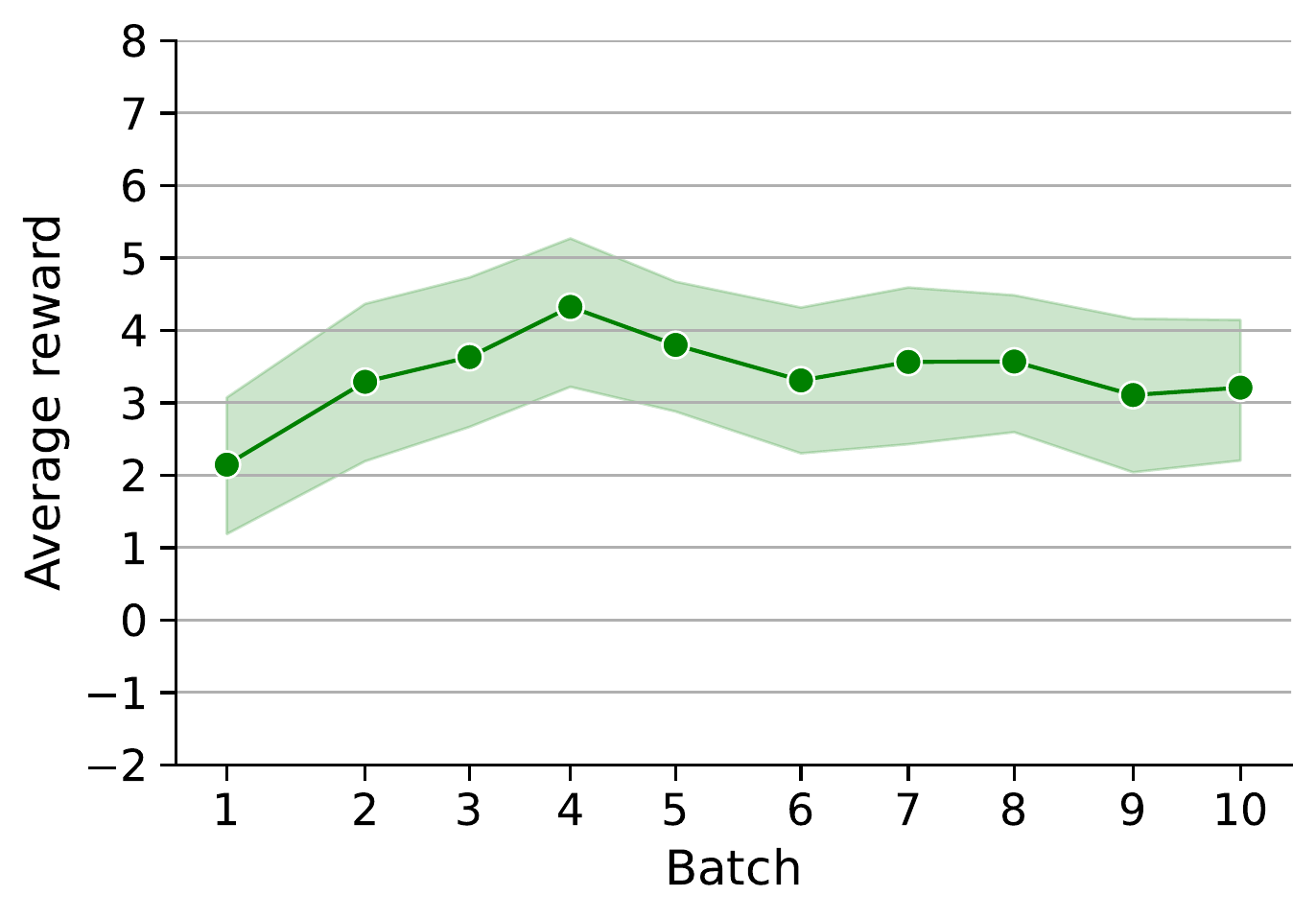}
      \includegraphics[width=.45\textwidth]{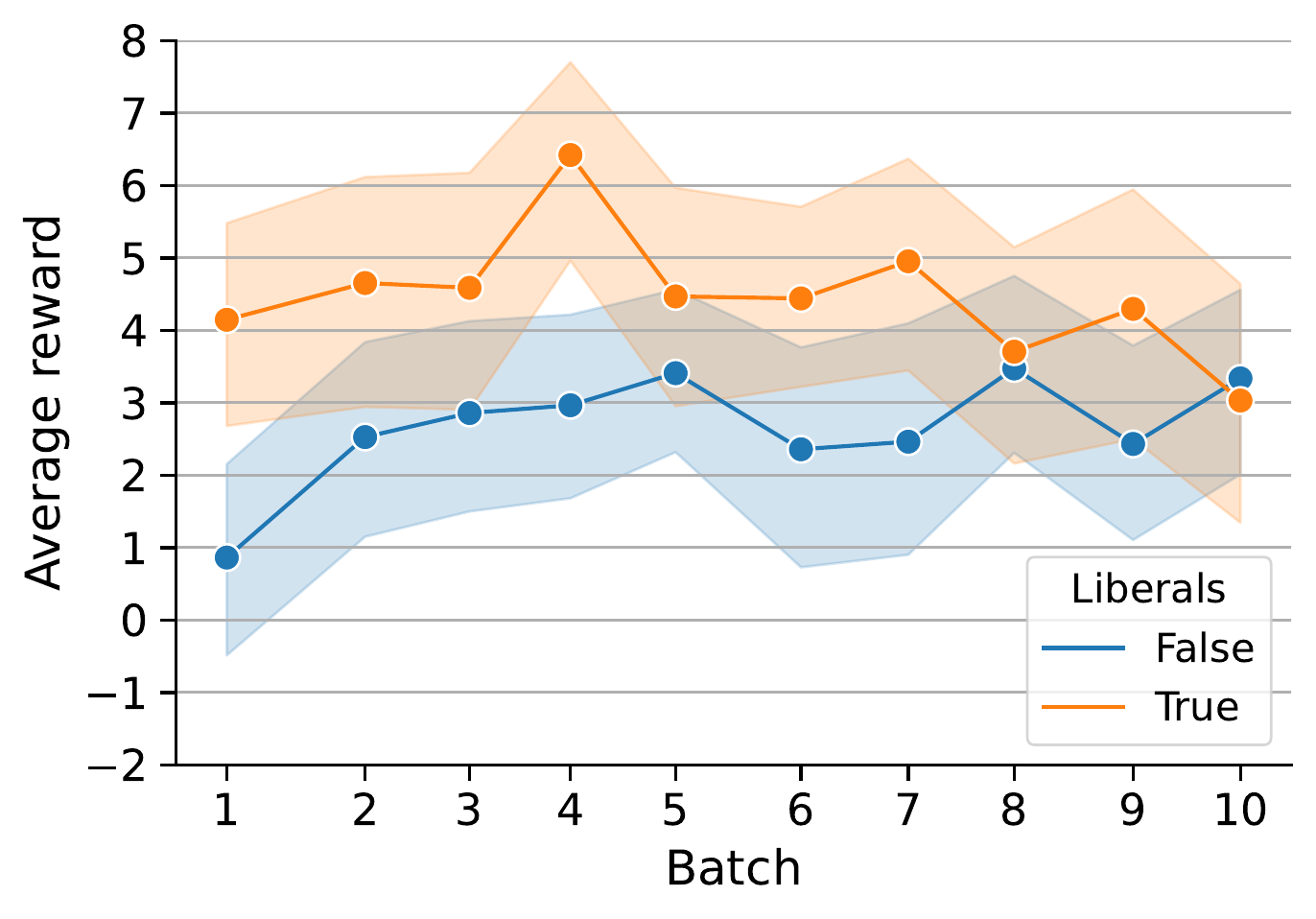}
        \includegraphics[width=.45\textwidth]{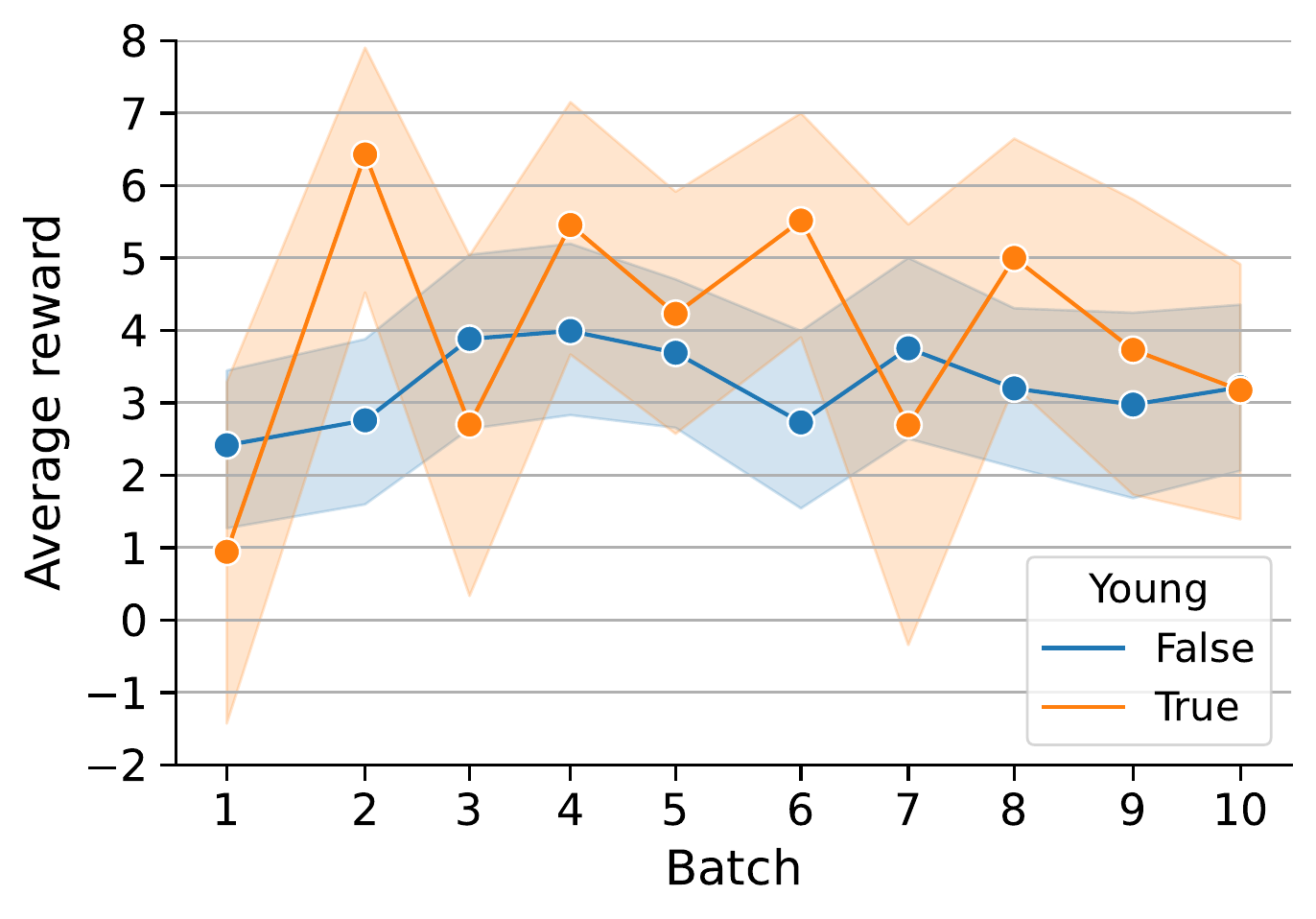}
    \includegraphics[width=.45\textwidth]{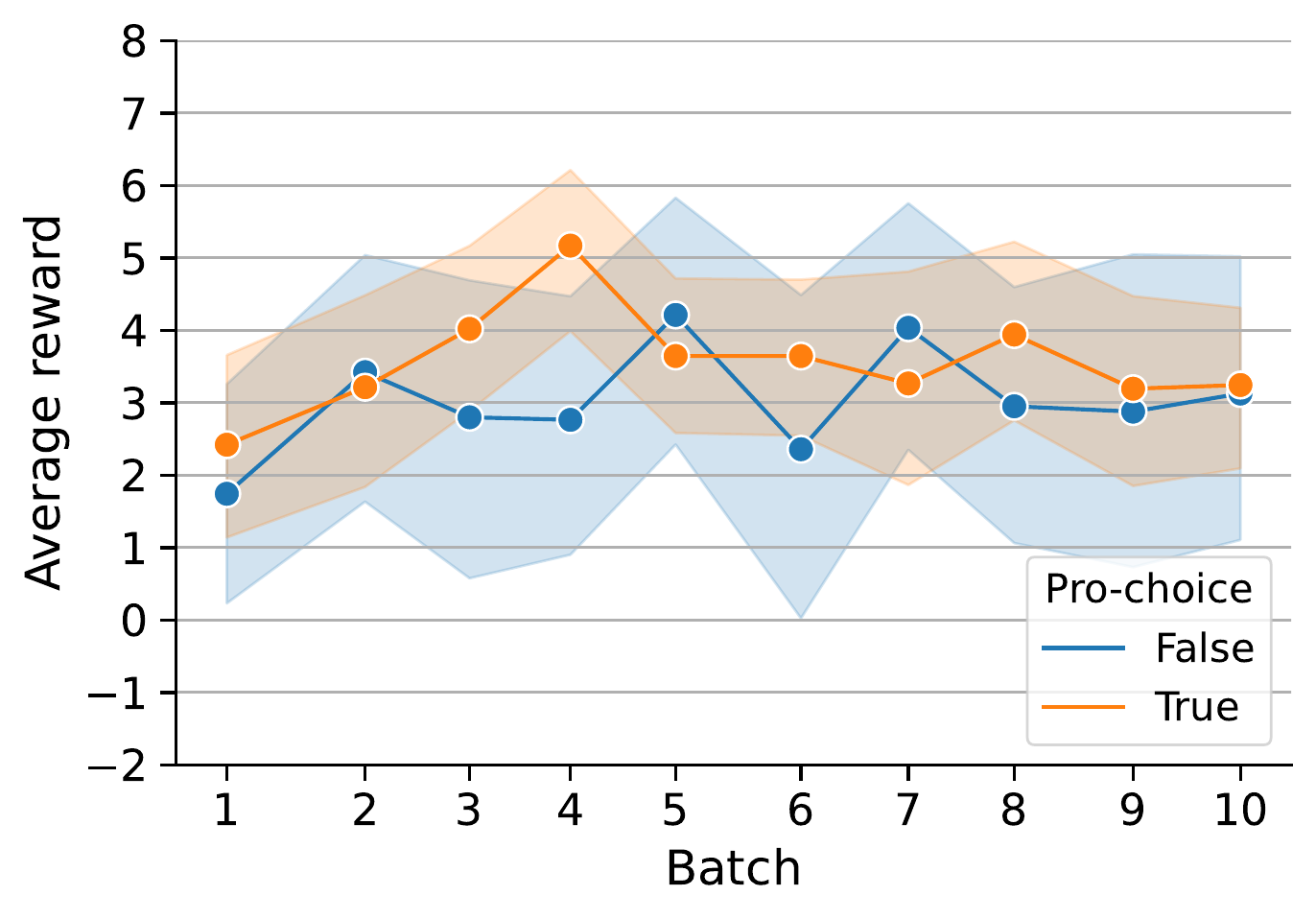}
    \caption{Evolution of average rewards over time. Error bars are 95\% confidence intervals around the sample average of reward within each batch.}
    \label{fig:avg_reward_evolution}
\end{figure}

Figure \ref{fig:avg_reward_evolution} shows average reward - the score assigned by respondents spanning extremely dissatisfied (-10) to extremely satisfied (10) - obtained over time. In the top left panel, we see that after batch 4, the average reward does not increase. We also plot the average reward obtained over time for different subgroups: ``liberals'' (political leaning $<4$ on a seven-point scale, with 1 being Strong Democrat) and ``conservatives'' (political leaning $\geq4$), ``young'' (age $<30$) and others, ``pro-choice'' (does not agree that abortion should be restricted, i.e., answered 1 to 3 on a five-point scale), and others (agree that abortion should be restricted). While the average reward mostly increases over time for ``conservatives,'' it starts high and ends up lower than where it started for ``liberals.'' For the other subgroups, the average reward oscillates a bit over time. However, for each of these subgroups except for liberals, the average reward is (weakly) higher than where it initially started, although the estimates are noisy.

\begin{figure}[H]
    \centering
    \includegraphics[width=.9\textwidth]{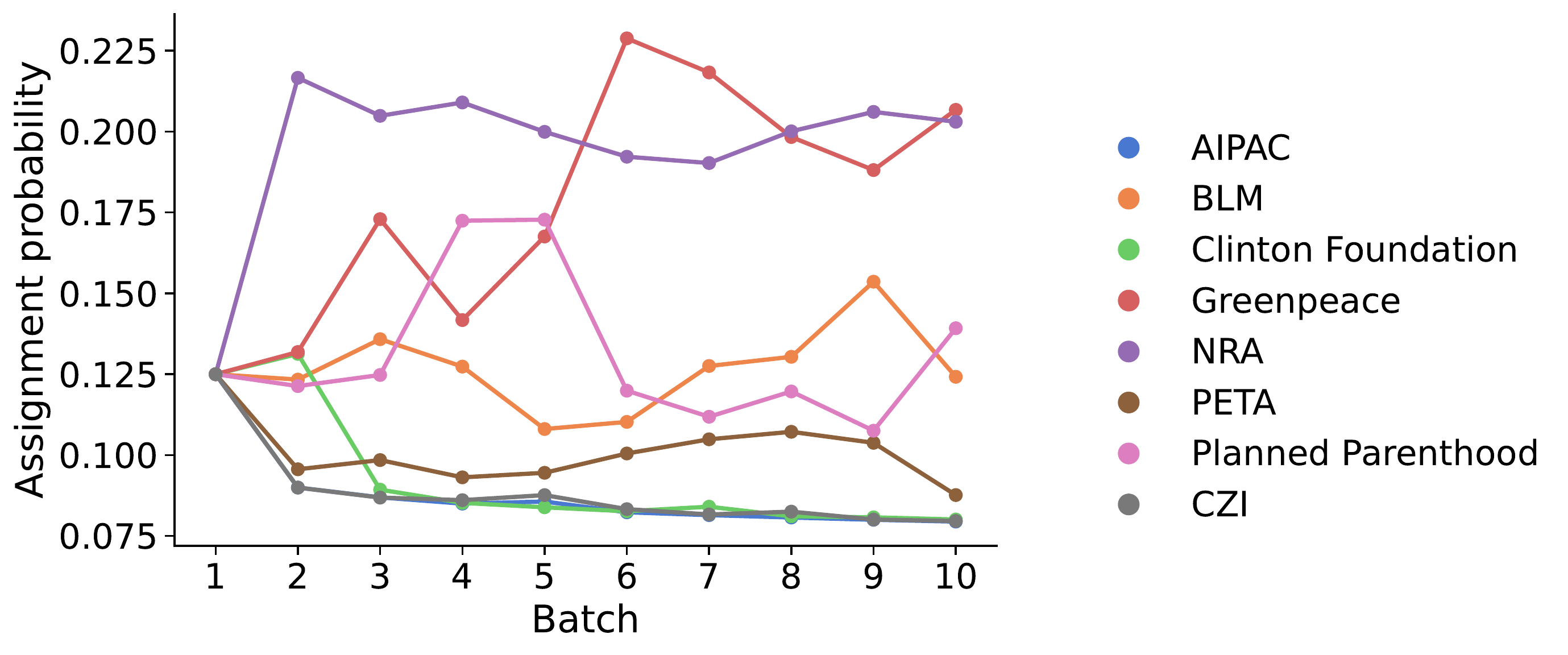}
    \caption{Average assignment probability in each batch}
    \label{fig:avg_arm_prob}
\end{figure}

Figure \ref{fig:avg_arm_prob} shows how average assignment probabilities evolve over time, i.e., $1/|B| \sum_{s \in B} e_{s}(X_s, w)$ for each batch $B$ and treatment $w$. This is a crude but informative measure to understand which arms are being favored by the algorithm. As we can see in Figure \ref{fig:avg_arm_prob}, the algorithm assigned NRA most often in the beginning of the experiment. The algorithm assigned Greenpeace more often later in the experiment - at the 6th batch; it is the most favored arm, and it continues to be either the first or the second most favored arm after that. Planned Parenthood was favored early on but declined in overall appeal relative to Greenpeace and finished with a third place. BLM had rather constant evolution over time and finished in a fourth place in terms of assignment probabilities.

%We note that after few observations the algorithm began assigning the Salvation army more often. However, Planned Parenthood was also favored early on, though later the algorithm preferred a mix of different charities, with the NRA taking second place by the end of the learning phase.

It is not possible to fully understand the behavior of the bandit by looking at average probabilities. An alternative approach is to determine a specific covariate value $x$ and observe the evolution of assignment probabilities for that value; i.e., $e_1(x, w), e_{51}(x, w), \cdots, e_{451}(x, w)$ for each treatment $w$.\footnote{Recall that assignment probabilities $e_t$ are fixed throughout for each $t$ in a batch of 50 observations, so that $e_1(x, w) = e_{2}(x, w) = \cdots = e_{50}(x, w)$ for each fixed context vector $x$ and arm $w$, and similar for other batches.} 

\begin{figure}[H]
    \centering
   % \hspace*{-2cm}
    \includegraphics[width=0.95\textwidth]{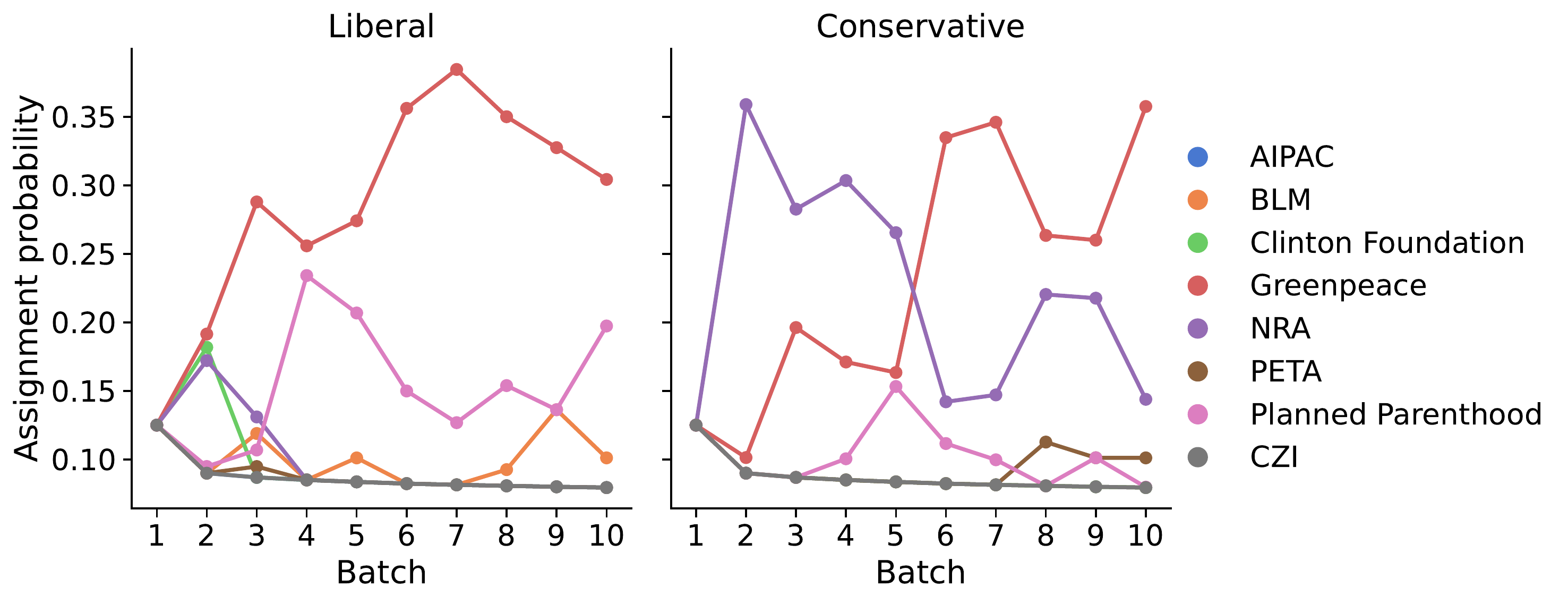}
    \caption{Evolution of assignment probabilities for the median participant according to political leaning. Each point is $e_t(x, w)$, where $x$ is the median covariate vectors among participants with the same political leaning.}
    \label{fig:assignment_prob_liberal}
\end{figure}

For example, what are the average assignment probabilities of each charity to participants who identify as liberal? To produce this, we first computed the median values of other covariates (e.g., views on abortion) among participants who reported their political leanings to be less than 4 (i.e., participants who identified as Strongly/Moderate/Leaning Democrat) and then computed the sequence of assignment probabilities for a hypothetical participant with these covariate values. Our results are in the first panel in Figure \ref{fig:assignment_prob_liberal}. Figure \ref{fig:assignment_prob_liberal} indicates that the assignment probabilities are different for ``liberals'' and ``conservatives.'' For liberals, Greenpeace is the most favored arm and Planned Parenthood is the second most favored arm by the end of the learning phase. For conservatives, NRA was favored early on, but Greenpeace takes over after the 5th batch.

\begin{figure}[H]
    \centering
   % \hspace*{-2cm}
    \includegraphics[width=0.95\textwidth]{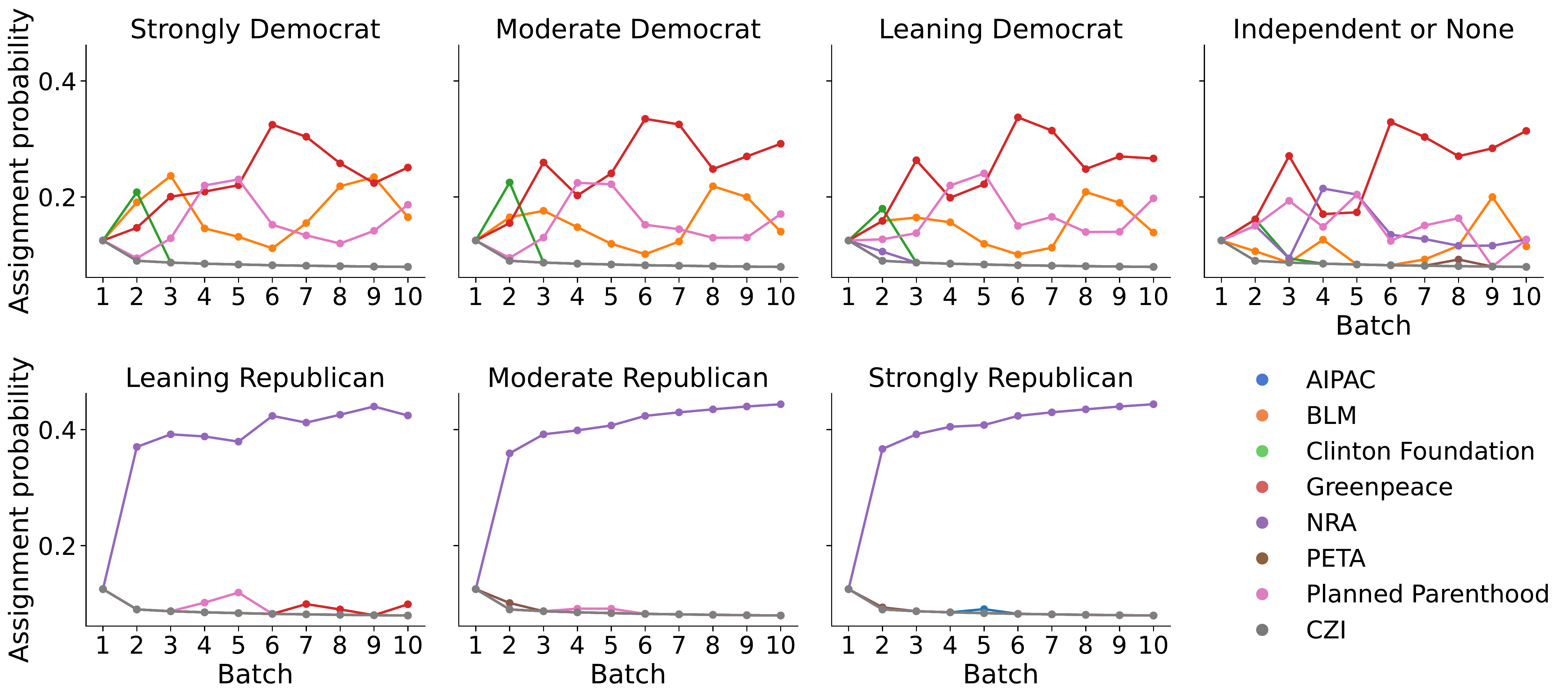}
    \caption{This figure shows the evolution of assignment probabilities for the median participant according to political leaning. Each point is $e_t(x, w)$, where $x$ is the median covariate vector among participants with the same political leaning.}
    \label{fig:assignment_prob_politics}
\end{figure}

We analogously create and plot the same figures across all political subgroups in Figure \ref{fig:assignment_prob_politics}. For Democrats and independent participants, Greenpeace is favored by the algorithm. On the other hand, for Republicans, the algorithm strongly favors NRA.

\begin{figure}[H]
    \centering
 %   \hspace*{-2cm}
    \includegraphics[width=0.95\textwidth]{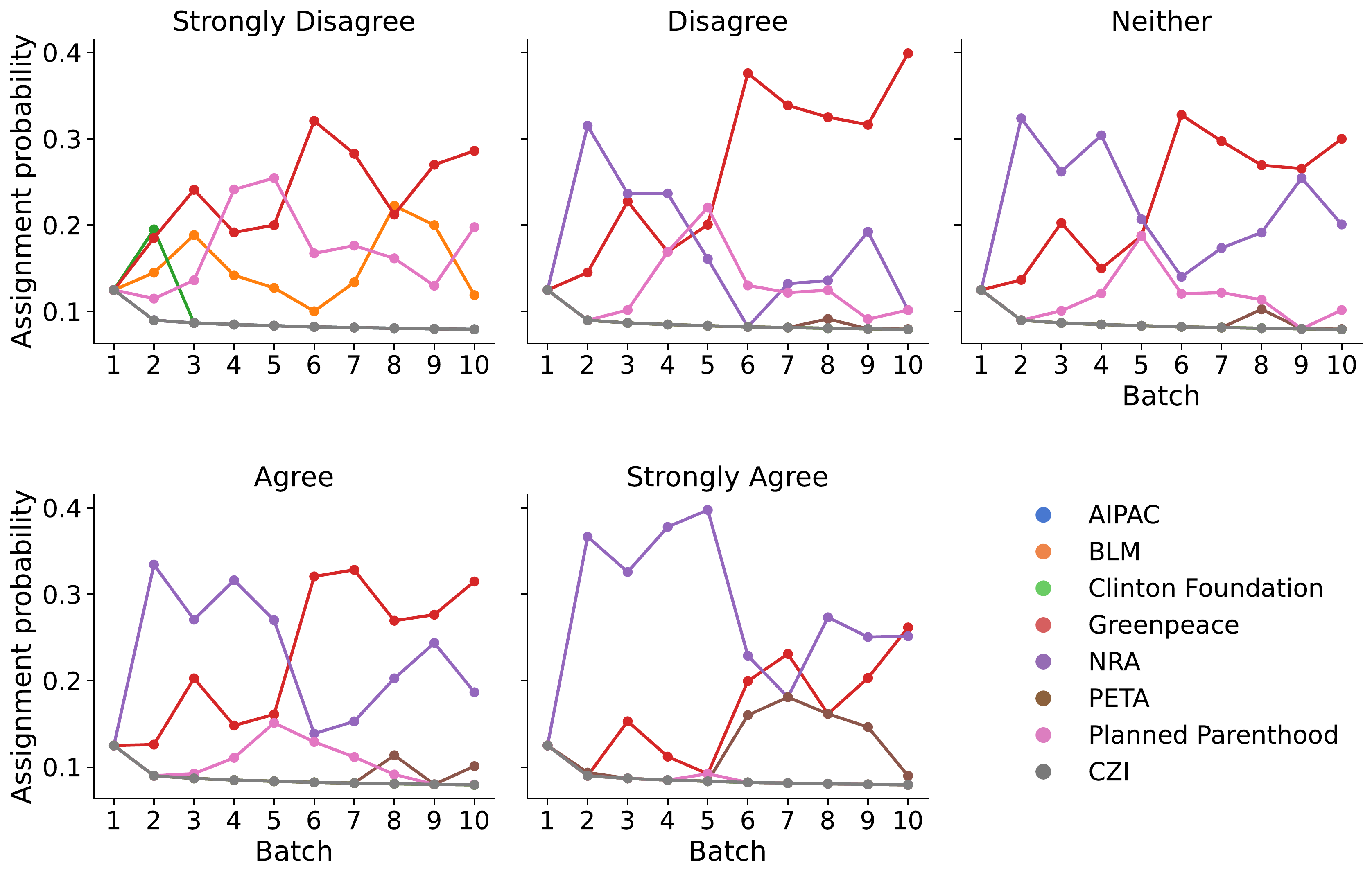}
    \caption{This figure shows the evolution of assignment probabilities for the median participant according to their views on abortion. The statement is ``Abortion should be banned or aggressively restricted.'' Each point is $e_t(x, w)$, where $x$ is the median covariate vector among participants with the same views on abortion.}
    \label{fig:assignment_prob_arms}
\end{figure}

Participants who favor abortion restrictions also have very different assignment probabilities compared to those who do not. Figure \ref{fig:assignment_prob_arms} indicates that among those who strongly agree with the statement ``Abortion should be banned or aggressively restricted,'' NRA is generally the most favored arm, while Greenpeace is generally the most favored arm for other subgroups.

\begin{figure}[H]
    \centering
    \includegraphics[width=0.9\textwidth]{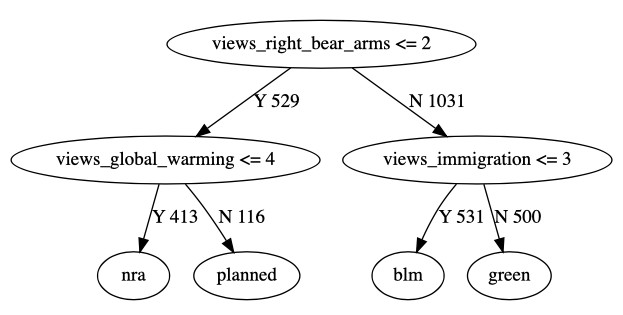}
    \caption{Learned contextual policy}
    \label{fig:learned_policy}
\end{figure}

\paragraph{Learned policies.} At the end of the learning phase, the four charities selected according to the frequency score (\ref{eq:freq}) were BLM, Planned Parenthood, Greenpeace, and NRA. We then used policyTree to estimate the best contextual policy using data from the learning phase, restricted to those four treatment arms. Figure \ref{fig:learned_policy} shows the contextual policy $\htree$ that was estimated using these four arms. The estimated best non-contextual policy $\hfixed$ assigns Greenpeace to every observation.

\begin{figure}
\centering
    \includegraphics[width=0.65\textwidth]{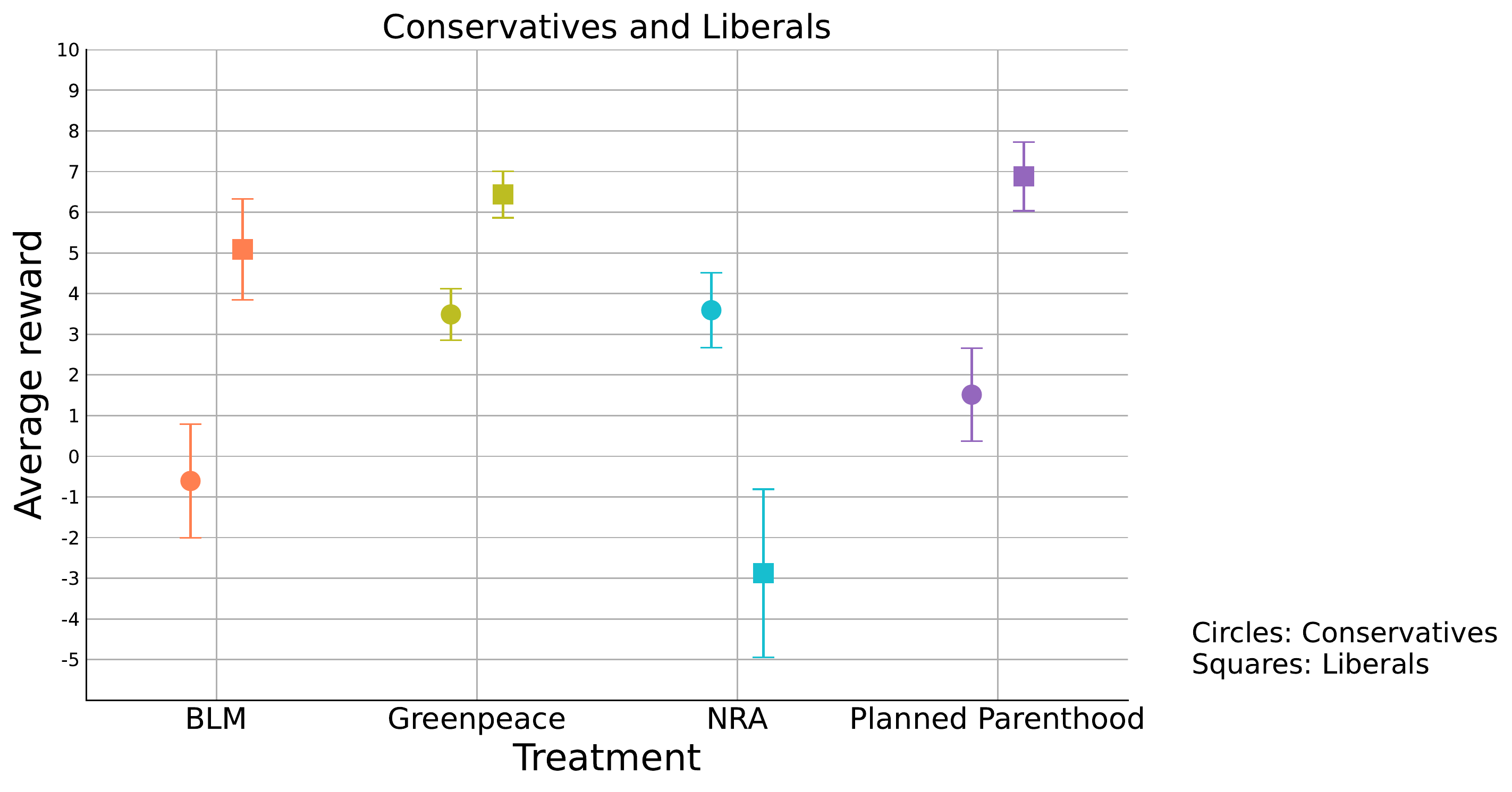}
    \includegraphics[width=0.7\textwidth]{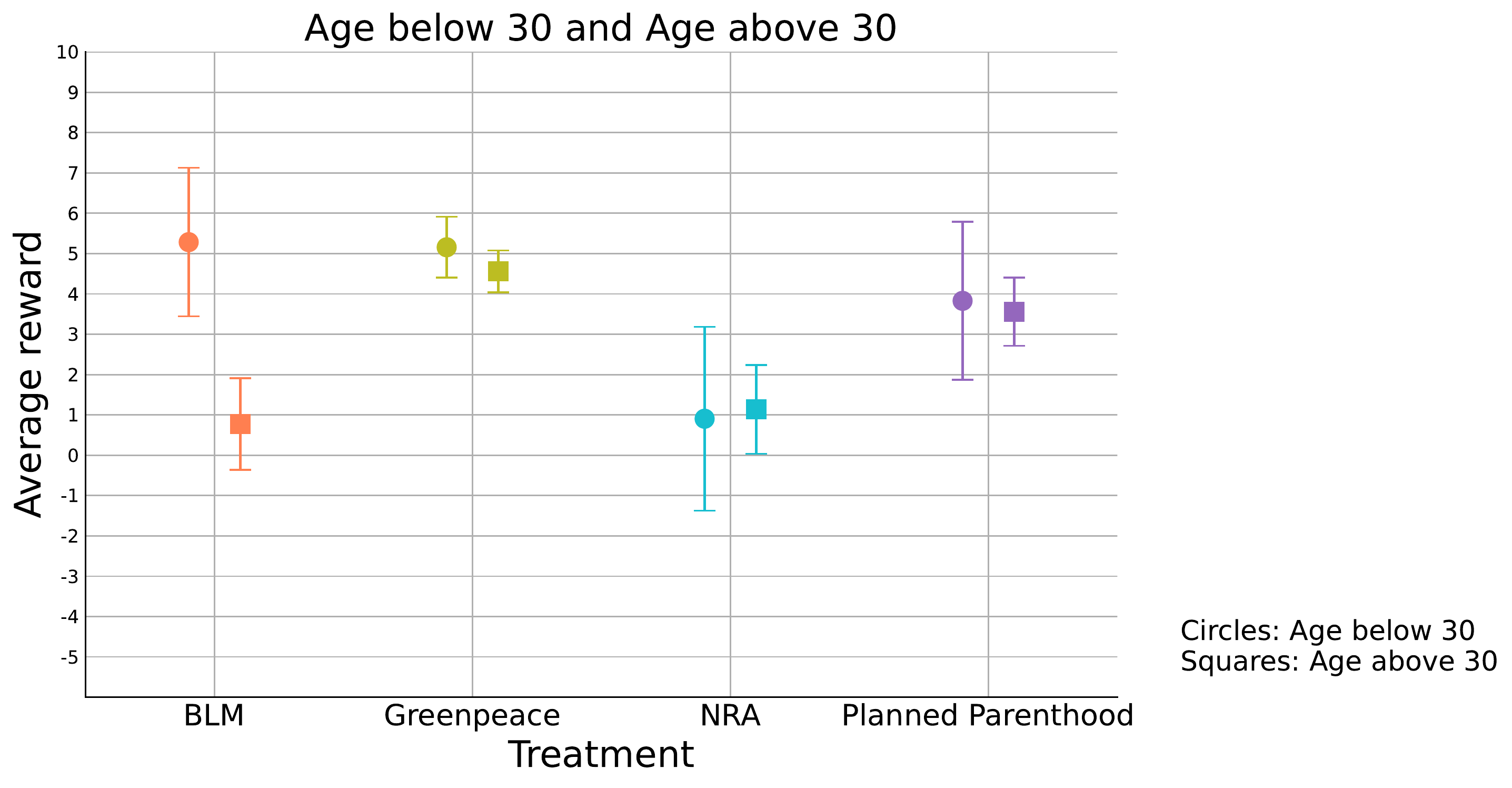}
    \includegraphics[width=0.7\textwidth]{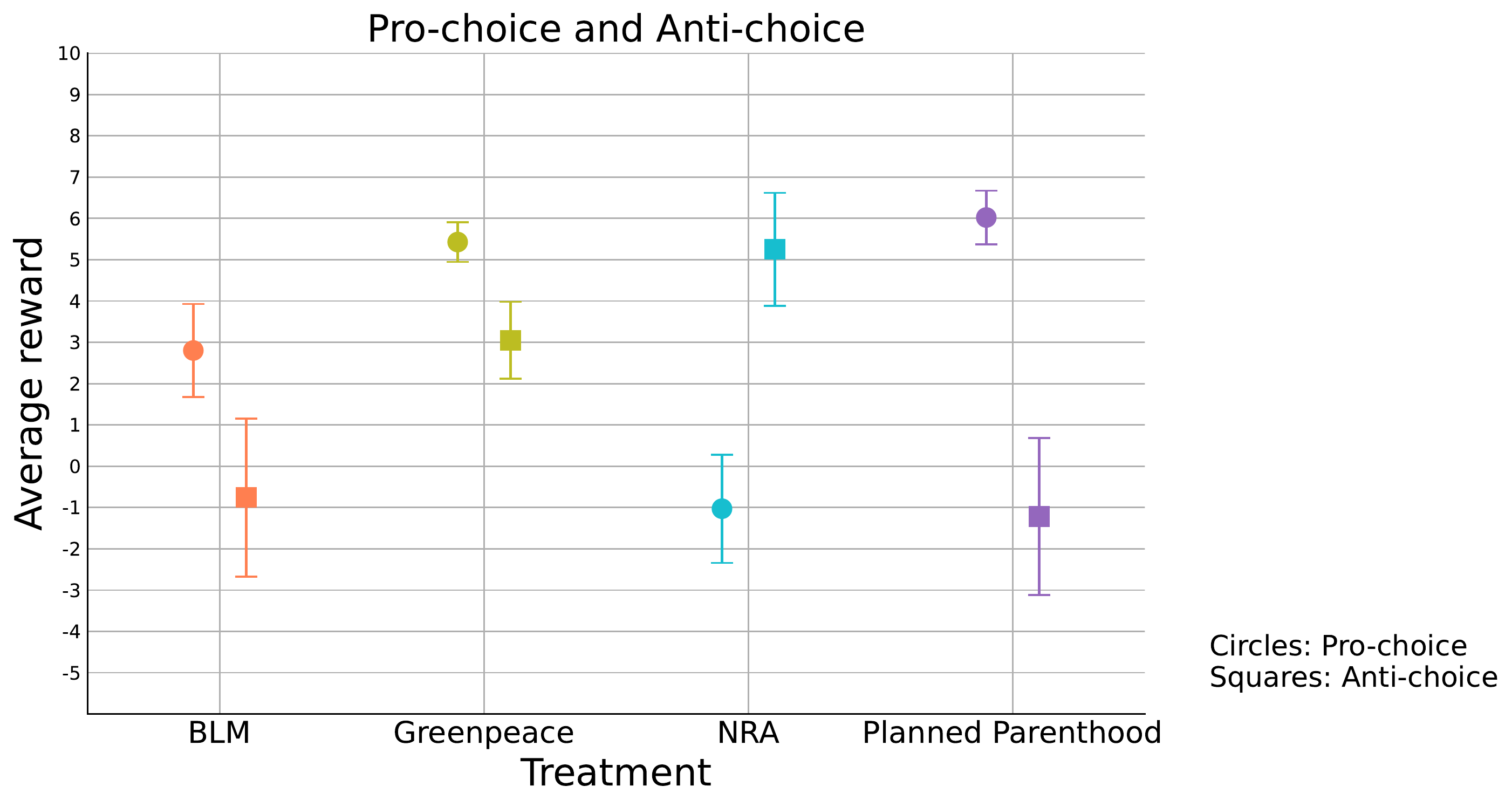}
    \caption{This figure shows average rewards and their 95\% confidence intervals for different subgroups. Means and standards errors that were used to produce this figure are reported in Table \ref{tab:subgroup_policy_values}. ``Liberals'' are those who are Democrat-leaning. ``Conservatives'' are Republican- and independent-leaning participants. ``Pro-choice'' participants are those who strongly disagree/somewhat disagree/neither agree nor disagree with the statement ``Abortion should be banned or aggressively restricted.'' The rewards are calculated using the full sample (i.e., learning and evaluation data).}
    \label{fig:subgroup_reward}
\end{figure}

\clearpage
\paragraph{Subgroups}  Table \ref{tab:subgroup_policy_values} shows the average rewards and their standard errors for different subgroups for the four selected arms which are BLM, Greenpeace, NRA and Planned Parenthood. The subgroups are the ones that appear in Figure \ref{fig:avg_reward_evolution}. This time, we use the entire sample (i.e., learning and evaluation data) to compute the average rewards. Figure \ref{fig:subgroup_reward} plots those means along with their 95\% confidence intervals. 

We observe variations across arms and across groups. When we classify participants based on their political leanings, we find that Planned Parenthood is the most favored charity among liberals and NRA is the most favored charity among conservatives. Point estimates indicate that the benefit of being assigned to the best charity over being assigned to Greenpeace is 7.2\% for liberals and 2.6\% for conservatives.

When classifying participants by age group, we find that those who are below age 30 favor BLM the most, and those who are above age 30 favor Greenpeace the most. Assigning BLM instead of Greenpeace to those below age 30 increases their outcome by 2\% in terms of point estimates.

When classifying participants by abortion views, we find that pro-choice participants favor Planned Parenthood the most and those who are anti-choice favor NRA the most. Assigning Planned Parenthood to pro-choice participants and NRA to anti-choice participants, instead of Greenpeace, increases their outcomes by 11\% and 71\%, respectively.

Some charities can be polarizing. For example, BLM is highly favored by liberals and those below age 30, but BLM is not favored by conservatives and those who are anti-choice. On the other hand, Greenpeace is less polarizing.

\begin{table}[H]
    \centering
    \begin{tabular}{lcccccccc}
\toprule
Subgroups & n &  Green &    BLM &    NRA &  Planned &  BLM- &  NRA- &  Planned-\\
 & & & & & & Green & Green & Green \\

\midrule
    Liberals &  1196 &  6.431 &  5.082 & -2.909 &    \textbf{6.895} &     -1.349 &     -9.340 &          0.464 \\
              &       &  (0.291) &  (0.634) &  (1.057) &    (0.430) &      (0.698) &      (1.097) &          (0.519) \\
              &\\
 Conservatives &  1869 &  3.493 & -0.622 &  \textbf{3.584} &    1.499 &     -4.116 &      0.091 &         -1.994 \\
              &       &  (0.323) &  (0.713) &  (0.472) &    (0.584) &  (0.782) & (0.572) & (0.667) \\
              & \\
      Age below 30 &   569 &  5.151 &  \textbf{5.255} &  0.913 &    3.846 &      0.104 &     -4.238 &         -1.305 \\
                &       &  (0.386) &  (0.940) &  (1.165) &    (0.987) &      (1.016) &      (1.227) &          (1.060) \\
                &\\ 
    Age above 30 &  2491 &  \textbf{4.565} &  0.767 &  1.115 &    3.549 &     -3.798 &     -3.450 &         -1.016 \\
               &       &  (0.265) &  (0.579) &  (0.562) &    (0.433) &      (0.636) &      (0.621) &          (0.507) \\
               &\\
    Pro-choice &  2042 &  5.427 &  2.797 & -1.050 &    \textbf{6.025} &     -2.630 &     -6.478 &          0.597 \\
               &       &  (0.245) &  (0.573) &  (0.669) &    (0.331) & (0.624) & (0.712) &  (0.412) \\
               &\\
  Anti-choice &  1022 &  3.059 & -0.784 &  \textbf{5.245} &   -1.233 &     -3.843 &      2.186 &         -4.292 \\
               &       &  (0.475) &  (0.976) &  (0.705) &    (0.971) &      (1.085) &    (0.850) &          (1.082) \\
\bottomrule
\end{tabular}

    \caption{This table shows the average reward for different subgroups. Standard errors are in parentheses. ``Liberals'' are those who are Democrat-leaning. ``Conservatives'' are Republican- and independent-leaning participants. ``Pro-choice'' participants are those who strongly disagree/somewhat disagree/neither agree nor disagree with the statement ``Abortion should be banned or aggressively restricted.'' The rewards are calculated using the full sample (i.e., learning and evaluation data).}
    \label{tab:subgroup_policy_values}
\end{table}

\clearpage
\paragraph{Evaluating Policies}

Let $\hpi$ be the policy learned at the end of the learning phase, and let $\pi^{N}$ denote the policy that always assigns the best (learned) fixed policy, learned using the data in the learning phase. The hypothesis we want to test is the following:
\begin{equation}
    H_0: \EE[Y_i(\htree(X_i))] \leq \EE[Y_i(\hfixed(X_i))].
\end{equation}

We use data from the evaluation phase to test this hypothesis. (Note that biases can arise if the same data is used to select a policy and evaluate it.) The estimated values of the learned contextual policy and the best fixed policy are given in Table \ref{tab:learned_policy_values}.  Column 3 in Table~\ref{tab:learned_policy_values} reports the difference between the value of the learned contextual policy and the best fixed policy, along with its standard error and p-value, validating our hypothesis as it shows that rewards are higher in the contextual policy setting compared to the best fixed policy. We also report the values of all fixed policies in Table \ref{tab:fixed_policy_values}.

\begin{table}[H]
    \centering
    \begin{tabular}{lrrrrr}
\toprule
{} &  Est. Value &  Std. Error & Est. Diff & Std. Error & p-value \\
Policy        &                 &             \\
\midrule
Best fixed policy (Greenpeace) &           4.687 &       0.208  \\
Learned contextual policy  &           5.653 &       0.216  &  0.966 &       0.300 &    0.001  \\
\bottomrule
\end{tabular}

% \begin{tabular}{lrrr}
% \toprule
% {} &  Learned contextual policy &  Best fixed policy & Contrast \\
% \midrule
% Value estimate  &           5.653 &       4.687 &  0.966  \\
% Std. Error &    0.216 &  0.208 &  0.300      \\
% p-value & - & - &  0.001
% \bottomrule
% \end{tabular}

    \caption{This table reports the value of learned contextual policy and the best fixed (or non-contextual) policy. Columns 1-2 show the estimate of the policy value  and its standard error, and columns 3-5 refer to the difference between the learned contextual policy and the best fixed policy. The values are calculated using evaluation data.}
    \label{tab:learned_policy_values}
\end{table}

\begin{table}[H]
    \centering
    \begin{tabular}{lrr}
\toprule
{} &  Est. Value &  Std. Error \\
Policy     &                 &             \\
\midrule
AIPAC         &           1.965 &       0.840 \\
BLM        &           1.181 &       0.939 \\
Clinton Foundation      &          -0.963 &       0.748 \\
Greenpeace       &           4.687 &       0.208 \\
NRA       &           0.960 &       1.088 \\
PETA      &           3.987 &       0.842 \\
Planned Parenthood    &           2.887 &       0.818 \\
Chan-Zuckerberg Initiative &           0.579 &       0.875 \\
\bottomrule
\end{tabular}
    \caption{Values of non-contextual policies. The values are calculated using evaluation data.}
    \label{tab:fixed_policy_values}
\end{table}

The learned policy divides the covariate space into (possibly empty) regions $R_1, ..., R_K$ defined by $R_w := \{ x \in \mathcal{X} : \htree(x) = w \}$. The next hypothesis we want to test is the following:

\begin{equation}
    H_0: \EE[Y_i(\htree(X_i))|R_w] \leq \EE[Y_i(\hfixed(X_i))|R_w].
\end{equation}

Table \ref{tab:contrast_estimates_per_region} shows the contrast estimates per region. For the region where NRA was recommended, we successfully validate this hypothesis. For the region where BLM was recommended, the difference seems to be too small to detect even with the evaluation data (which is designed to detect these differences). Unfortunately, for the region where Planned Parenthood was recommended, it appears that Greenpeace dominates. However, only a few users fall into this region.

\begin{table}[H]
    \centering
    \begin{tabular}{lrrrr}
\toprule
{} &  Est. Diff &  Std. Error &  p-value &  n \\
Contrast        &        &             &          &       \\
\midrule
BLM - Greenpeace    & -0.309 &       0.305 &    0.155 &   461 \\
NRA - Greenpeace     &  4.264 &       0.340 &    0.000 &   437 \\
Planned Parenthood - Greenpeace & -2.137 &       0.346 &    0.000 &   125 \\
\bottomrule
\end{tabular}

    \caption{Contrast estimates per region. The values are calculated using evaluation data.}
    \label{tab:contrast_estimates_per_region}
\end{table}

% Finally, Table \ref{tab:average_covariate_per_region} shows estimates of the expected covariate value per region.

% \begin{table}[H]
%     \centering
%     \input{tables/average_covariate_per_region}
%     \caption{Estimate of average covariate value per learned contextual policy region.}
%     \label{tab:average_covariate_per_region}
% \end{table}

\section{Bandits versus Uniform Randomization}
\label{sec:FinalSimul}
In this section, we use the additional data collected from the main experiment to reevaluate the benefits of the bandit design over standard uniform sampling. Ideally, we would run our bandit design and uniform sampling several times and then compare the value of the policy learned and cumulative regret. However, this would be prohibitively expensive, so we instead consider two approaches to evaluating the benefits.

\begin{figure}[H]
    \centering
    \includegraphics[width=0.6\textwidth]{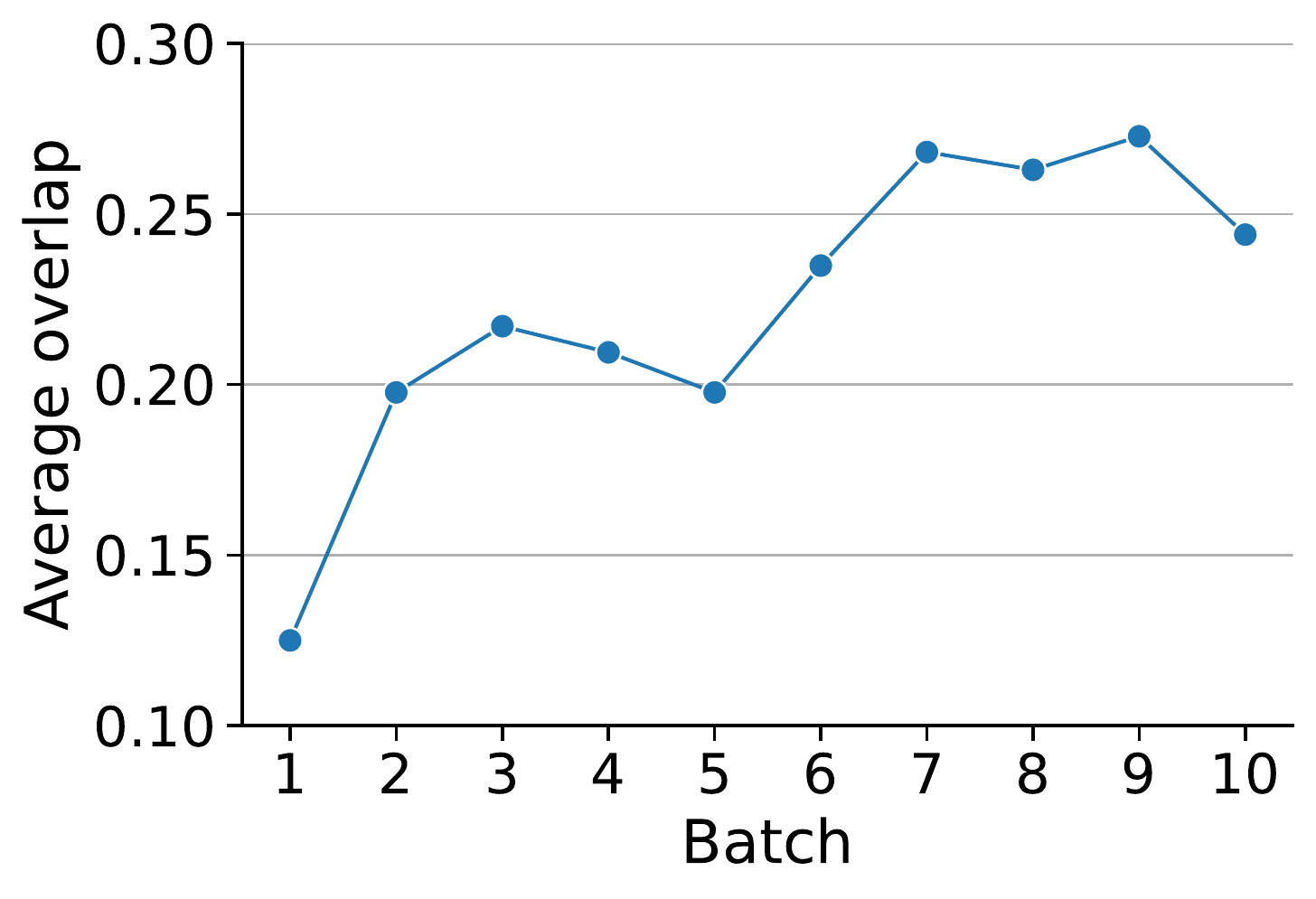}
    \caption{This figure shows the evolution of average probability for choosing the arm recommended by the learned policy $\hat\pi^C$ (learned at the end of the learning phase of the main experiment) in every batch of the main experiment.}
    \label{fig:overlap}
\end{figure}

\paragraph{Summary statistics from the main experiment highlight adaptivity benefits.} Contextual bandits can outperform uniform sampling for policy learning when they collect more data on potentially optimal policies, since this enables more accurate selection of the best policy. To evaluate whether this benefit was present in the main experiment, we present evidence that indeed, the learning phase collected more data about our ultimately selected policy than would have been collected under uniform randomized sampling (used in the first batch). From Figure~\ref{fig:overlap}, we observe that the average probability for choosing the arm recommended by the learned policy (empirically optimal policy) increases over time. This data collection allowed us to understand with higher precision at the end of the learning phase that this policy was indeed a high-performing policy. Further, from Table~\ref{tab:average_reward}, we see that the average reward of the TreeBagging bandit we use is higher than the estimated reward from uniform assignment, providing evidence that our algorithm did indeed achieve lower cumulative regret.

\begin{table}[H]
    \centering
    \begin{tabular}{lrr}
\toprule
{} &  Mean &  Std.Error \\
Bandit          &                 &            \\
\midrule
TreeBagging(50) &           3.365 &      0.166 \\
Uniform & 1.913 & 0.831\\
\bottomrule
\end{tabular}

    \caption{Average rewards. The average reward of the TreeBagging(50) bandits is computed using data collected in the learning phase. The average reward of Uniform sampling is obtained by averaging the estimates of fixed policies in Table~\ref{tab:fixed_policy_values}, using data from the evaluation phase.}
    \label{tab:average_reward}
\end{table}

\paragraph{Simulation study based on main experiment data.} Our next exercise to evaluate the performance of our selected adaptive experimental design is to conduct a final simulation study designed to mirror our setting. Note that our earlier simulation study was based on pilot data. In this final simulation study we run our bandit design and uniform sampling as well as the other algorithms we considered in Section \ref{sec:parameter} several times on data generating processes learned from data collected in our learning and evaluation phases -- then compare the value of the policy learned and cumulative regret.

\begin{figure}[H]
    \centering
    \includegraphics[width=0.85\textwidth]{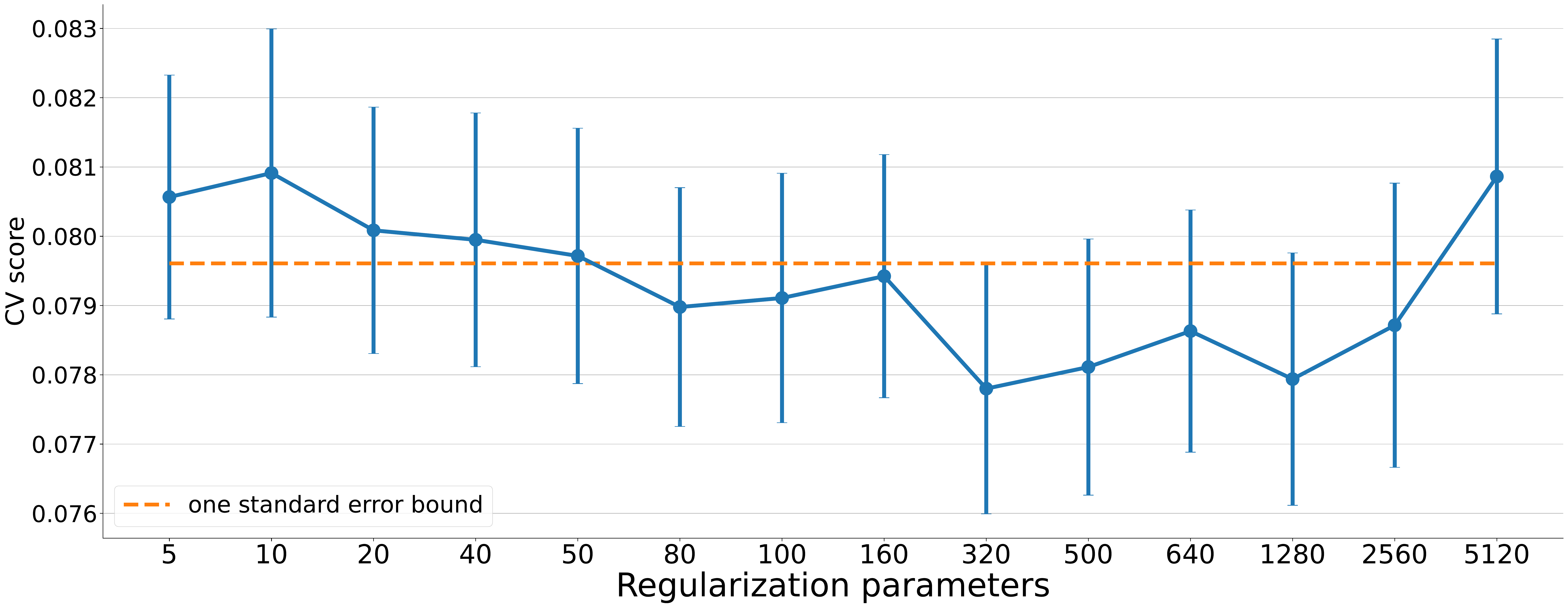}
    \caption{This figure displays normalized 10-fold cross-validation (CV) scores (average mean squared errors) for all regularization parameters. Error bars represent one standard error. The dotted line is the one standard error upper-bound of the smallest CV score (which is at 320).
    The parameters $\{80, 100, 160, 320, 500, 640, 1280, 2560\}$ are selected as their CV scores lie below this bound. }
    \label{fig:onese}
\end{figure}

The simulation design is the same as in Section \ref{sec:parameter} except that we use different sets of regularization parameters that best capture the heterogeneity in the main experiment data. Specifically, the set of regularization parameters (the factor multiplying the L2-norm of coefficients in the model described in Section \ref{sec:parameter}) are selected via the one standard error rule from the following list of regularization parameters, $\{5, 10, 20, 40, 50, 80, 100, 160, 320, 500, 640, 1280, 2560, 5120\}$. Figure~\ref{fig:onese} plots normalized 10-fold cross-validation (CV) scores for this list of regularization parameter. The responses and predicted values lie in the range $[-10,10]$, and before computing the mean squared errors, we normalize these to lie in $[0,1]$. Using the one standard error rule, which selects the regularization parameters with CV scores within a one standard error of the best CV score, we get the following list of regularization parameters $\{80, 100, 160, 320, 500, 640, 1280, 2560\}$.

Tables \ref{tab:post_simulation_learned_policy_value} and \ref{tab:post_simulation_regret} quantify the benefits of running our adaptive experiment design. Table \ref{tab:post_simulation_learned_policy_value} shows that the learned policy is at least as good as the one learned when collecting data uniformly, and, in fact, it improves upon it by a modest margin. Table \ref{tab:post_simulation_regret} shows that we indeed get modest benefits (about 20\% reduction) in terms of regret. 

\begin{table}[H]
    \centering
    \begin{tabular}{lrrrrrrrr}
\toprule
Regularization &  80 &  100 &  160 & 320 &  500 & 640 &  1280 & 2560 \\
Bandit & \\
\midrule
Uniform                &   5.804 &    5.785 &    5.764 &    5.706 &    5.619 &    5.621 &     5.504 &     5.328 \\
          &   (0.020) &    (0.018) &    (0.018) &    (0.017) &    (0.017) &    (0.016) &     (0.017) &     (0.019) \\

TreeBagging(50)       &   5.840 &    5.835 &    5.806 &    5.754 &    5.683 &    5.674 &     5.546 &     5.419 \\
 &   (0.019) &    (0.016) &    (0.017) &    (0.015) &    (0.015) &    (0.013) &     (0.015) &     (0.015) \\

BootstrapThompson     &   5.790 &    5.747 &    5.755 &    5.678 &    5.607 &    5.617 &     5.521 &     5.392 \\
 &   (0.021) &    (0.016) &    (0.018) &    (0.019)  &   (0.015) &    (0.017) & (0.015) &  (0.016) \\

BootstrapES            &   5.806 &    5.774 &    5.744 &    5.693 &    5.629 &    5.652 &     5.500 &     5.395 \\
     &   (0.020) &    (0.018) &    (0.018) &    (0.017) &    (0.016) &    (0.014) &     (0.019) &     (0.018) \\
BootstrapTTTS          &   5.779 &    5.773 &    5.769 &    5.696 &    5.633 &    5.642 &     5.534 &     5.379 \\
   &   (0.021) &    (0.018) &    (0.017) &  (0.015) &  (0.016) &    (0.016) & (0.017)  & (0.017) \\

Improvement & 100.63\% &  100.87\% &  100.73\% &  100.84\% &  101.13
\% &  100.94\%&   100.76\% &   101.70\% \\
\multicolumn{9}{l}{(TreeBagging(50) as \% of Uniform)}\\ 
\bottomrule
\end{tabular}
    \caption{This table provides the value of learned tree policy $\htree$, where the policy is learned from data collected using a bandit algorithm with the default parameters in Table \ref{tab:parameters}. Each column corresponds to a regularization parameter used in fitting the outcome model that underlies our simulation of participant outcomes. Lower regularization parameters correspond to higher heterogeneity in the simulated data. Averages and standard errors (in parentheses) are computed across over 1,000 simulations per cell. The last row indicates the improvement of TreeBagging(50) over Uniform in terms of the value of policy learned.}
    \label{tab:post_simulation_learned_policy_value}
\end{table}

\begin{table}[H]
    \centering
    \begin{tabular}{lrrrrrrrr}
\toprule
Regularization &  80 &  100 &  160 &  320 &  500 &  640 &  1280 &  2560 \\
Bandit & \\
\midrule
Uniform                &   4.611 &    4.556 &    4.475 &    4.309 &    4.156 &    4.105 &     3.811 &     3.500 \\
          &   (0.015) &    (0.011) &    (0.012) &    (0.011) &    (0.011) &   (0.012) &  (0.011) &  (0.011) \\
TreeBagging(50)       &   3.683 &    3.610 &    3.548 &    3.415 &    3.301 &    3.249 &     3.014 &     2.775 \\
  &   (0.012) &    (0.009) &    (0.011) &    (0.010) &   (0.010) &    (0.010) &  (0.009) &   (0.010) \\
BootstrapThompson      &   3.523 &    3.466 &    3.411 &    3.290 &    3.172 &    3.127 &     2.921 &     2.700 \\
 &   (0.012) &    (0.010) &    (0.010) &    (0.009) &    (0.009) &   (0.009) &   (0.009) &   (0.010) \\

BootstrapES          &   3.634 &    3.593 &    3.537 &    3.406 &    3.288 &    3.261 &     3.039 &     2.817 \\
     &   (0.012) &   (0.010) &    (0.009) &   (0.009) &   (0.009) &   (0.010) &   (0.009) &   (0.009) \\
BootstrapTTTS          &   3.636 &    3.603 &    3.531 &    3.412 &    3.286 &    3.256 &     3.031 &     2.811 \\
  &   (0.012) &    (0.009) &   (0.009) &    (0.010) &   (0.010) &   (0.009) &    (0.010) &    (0.009) \\

Reduction                           &  79.87\% &   79.24\% &   79.28\% &   79.24\% &   79.43\% &   79.15\% &    79.10\% &    79.30\% \\
\multicolumn{9}{l}{(TreeBagging(50) as \% of Uniform)}\\
\bottomrule
\end{tabular}

    \caption{This table shows the averaged per-period regret attained when data is collected via a bandit algorithm (indicated in the row labels) using default parameters in Table \ref{tab:parameters}. Each column indicates the regularization parameter used in fitting the outcome model that underlies our simulation of participant outcomes. Lower regularization parameters correspond to higher heterogeneity in the simulated data. Averages and standard errors (in parentheses) are computed across over 1,000 simulations per cell. The last row indicates the reduction in average regret of TreeBagging(50) over Uniform.
    }
    \label{tab:post_simulation_regret}
\end{table}

%Among the regularization parameters that we used in the simulation above, the parameter that corresponds to low heterogeneity best fits our data. To do that, we use 3-fold cross validation on the entire sample and choose the parameter that generates the lowest mean squared error. 

\subsection{Simulation Study Based on Pooled Data (Pilot and Main Experiment Data)}

\begin{figure}[H]
    \centering
    \includegraphics[width=0.85\textwidth]{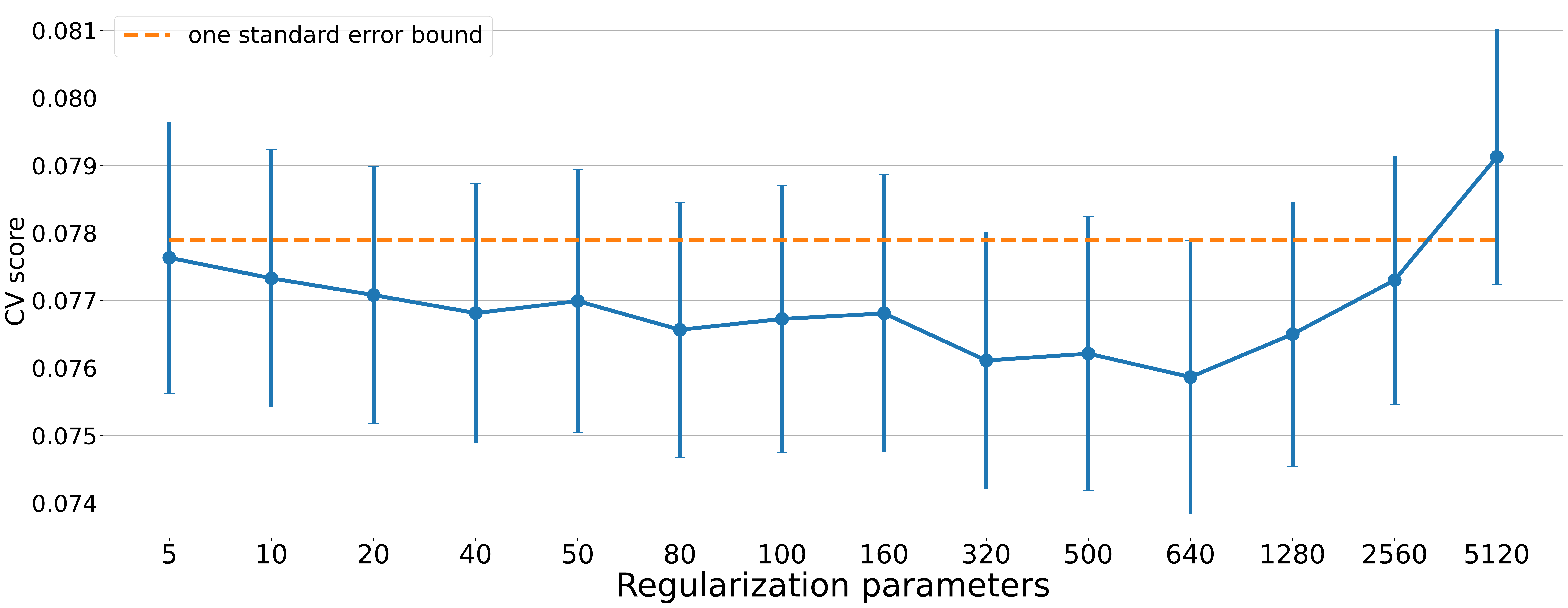}
    \caption{Normalized 10-fold cross-validation (CV) scores (average mean squared errors) for all regularization parameters. Error bars represent one standard error. The dotted line is the one standard error upper-bound of the smallest CV score (which is at 640).
    The parameters $\{5, 10, 20, 40, 50, 80, 100, 160, 320, 500, 640, 1280, 2560\}$ are selected as their CV scores lie below this bound. }
    \label{fig:onese_pooled}
\end{figure}

\begin{table}[H]
    \centering

  \begin{tabular}{lrrrrrrrrr}
\toprule
Regularization Parameter Value &   5 &   10 &   20 &   40 &   50 &   80 &   100 &   160 &   320  \\
\midrule
Uniform               &   5.88 &    5.85 &    5.86 &    5.84 &    5.84 &    5.84 &     5.83 &     5.79 &     5.76  \\
         &   0.02 &    0.02 &    0.02 &    0.02 &    0.02 &    0.02 &     0.02 &     0.02 &     0.02 \\
TreeBagging(50)    &   5.92 &    5.89 &    5.90 &    5.89 &    5.92 &    5.89 &     5.89 &     5.84 &     5.80  \\
 &   0.02 &    0.02 &    0.02 &    0.02 &    0.02 &    0.02 &     0.02 &     0.02 &     0.02 \\
 BootstrapThompson    &   5.83 &    5.83 &    5.84 &    5.82 &    5.81 &    5.82 &     5.81 &     5.79 &     5.75  \\
&   0.02 &    0.02 &    0.02 &    0.02 &    0.02 &    0.02 &     0.02 &     0.02 &     0.02  \\

BootstrapES            &   5.83 &    5.83 &    5.85 &    5.79 &    5.82 &    5.81 &     5.81 &     5.77 &     5.74 \\
      &   0.02 &    0.02 &    0.02 &    0.02 &    0.02 &    0.02 &     0.02 &     0.02 &     0.02\\
BootstrapTTTS        &   5.83 &    5.86 &    5.83 &    5.81 &    5.82 &    5.83 &     5.83 &     5.81 &     5.74 \\
   &   0.02 &    0.02 &    0.02 &    0.02 &    0.02 &    0.02 &     0.02 &     0.02 &     0.02 \\

Improvement                          & 100.76 &  100.78 &  100.60 &  100.82 &  101.34 &  100.79 &   101.00 &   100.90 &   100.67 \\
\multicolumn{10}{l}{(TreeBagging(50) as \% of Uniform)}\\ 
\bottomrule
\end{tabular}

    \caption{This table shows the value of learned tree policy $\htree$ when data is collected using bandit algorithms (indicated in rows) with the default parameters in Table \ref{tab:parameters}. Each column corresponds to a regularization parameter used in fitting the outcome model that underlies our simulation of participant outcomes. Lower regularization parameters correspond to higher heterogeneity in the simulated data. Averages and standard errors (in parentheses) are computed across over 1,000 simulations per cell. The last row indicates the improvement of TreeBagging(50) over Uniform in terms of the value of policy learned.
    }
\end{table}

\begin{table}[H]
    \centering

  \begin{tabular}{lrrrr}
\toprule
Regularization &     500 &   640 &   1280 &   2560 \\
\midrule
Uniform               &      5.74 &     5.69 &      5.61 &      5.52 \\
         &     0.02 &     0.02 &      0.02 &      0.02 \\
TreeBagging(50)    &       5.75 &     5.74 &      5.69 &      5.55 \\
 &      0.02 &     0.02 &      0.01 &      0.02 \\
 BootstrapThompson    &       5.71 &     5.69 &      5.60 &      5.51 \\
&      0.02 &     0.02 &      0.02 &      0.02 \\

BootstrapES            &       5.73 &     5.69 &      5.58 &      5.51 \\
      &      0.02 &     0.02 &      0.02 &      0.02 \\
BootstrapTTTS        &    5.75 &     5.68 &      5.61 &      5.49 \\
   &      0.02 &     0.02 &      0.02 &      0.02 \\

Improvement                          &  100.13 &   100.77 &    101.42 &    100.67 \\
\multicolumn{5}{l}{(TreeBagging(50) as \% of Uniform)}\\ 
\bottomrule
\end{tabular}

    \caption{This table shows the value of learned tree policy $\htree$ when data is collected using bandit algorithms (indicated in rows) with the default parameters in Table \ref{tab:parameters}. Each column corresponds to a regularization parameter used in fitting the outcome model that underlies our simulation of participant outcomes. Lower regularization parameters correspond to higher heterogeneity in the simulated data. Averages and standard errors (in parentheses) are computed across over 1,000 simulations per cell. The last row indicates the improvement of TreeBagging(50) over Uniform in terms of the value of policy learned.
    }
\end{table}

\begin{table}[H]
    \centering
   \begin{tabular}{lrrrrrrrrr}
\toprule
Regularization &   5 &   10 &   20 &   40 &   50 &   80 &   100 &   160 &   320\\
\midrule
Uniform             &   4.83 &    4.77 &    4.76 &    4.75 &    4.73 &    4.67 &     4.65 &     4.57 &     4.48  \\
        &   0.01 &    0.01 &    0.01 &    0.01 &    0.01 &    0.01 &     0.01 &     0.01 &     0.01  \\
TreeBagging(50)       &   3.84 &    3.80 &    3.78 &    3.75 &    3.72 &    3.69 &     3.67 &     3.61 &     3.55  \\
  &   0.01 &    0.01 &    0.01 &    0.01 &    0.01 &    0.01 &     0.01 &     0.01 &     0.01 \\
BootstrapThompson     &   3.67 &    3.64 &    3.61 &    3.60 &    3.58 &    3.55 &     3.53 &     3.47 &     3.38 \\
&   0.01 &    0.01 &    0.01 &    0.01 &    0.01 &    0.01 &     0.01 &     0.01 &     0.01  \\

BootstrapES           &   3.80 &    3.75 &    3.76 &    3.73 &    3.71 &    3.67 &     3.65 &     3.59 &     3.53  \\
     &   0.01 &    0.01 &    0.01 &    0.01 &    0.01 &    0.01 &     0.01 &     0.01 &     0.01 \\
BootstrapTTTS         &   3.79 &    3.75 &    3.75 &    3.73 &    3.71 &    3.67 &     3.64 &     3.57 &     3.52  \\
    &   0.01 &    0.01 &    0.01 &    0.01 &    0.01 &    0.01 &     0.01 &     0.01 &     0.01 \\

Reduction                           &  79.41 &   79.72 &   79.49 &   78.96 &   78.82 &   79.01 &    78.96 &    78.89 &    79.16 \\
\multicolumn{10}{l}{(TreeBagging(50) as \% of Uniform)}\\ 
\bottomrule
\end{tabular}

    \caption{This table displays the averaged per-period regret attained when data is collected using bandit algorithms (indicated in rows) with default parameters in Table \ref{tab:parameters}. Each column indicates the regularization parameter used in fitting the outcome model that underlies our simulation of participant outcomes. Lower regularization parameters correspond to higher heterogeneity in the simulated data. Averages and standard errors (in parentheses) are computed across over 1,000 simulations per cell. The last row indicates the reduction in average regret of TreeBagging(50) over Uniform.
    }
\end{table}

\begin{table}[H]
    \centering
   \begin{tabular}{lrrrr}
\toprule
Regularization &   500 &   640 &   1280 &   2560 \\
\midrule
Uniform             &      4.37 &     4.31 &      4.09 &      3.80 \\
        &     0.01 &     0.01 &      0.01 &      0.01 \\
TreeBagging(50)       &      3.44 &     3.39 &      3.21 &      2.99 \\
  &    0.01 &     0.01 &      0.01 &      0.01 \\
BootstrapThompson     &      3.31 &     3.27 &      3.10 &      2.92 \\
&       0.01 &     0.01 &      0.01 &      0.01 \\

BootstrapES           &       3.45 &     3.40 &      3.23 &      3.04 \\
     &     0.01 &     0.01 &      0.01 &      0.01 \\
BootstrapTTTS         &     3.44 &     3.39 &      3.23 &      3.04 \\
    &    0.01 &     0.01 &      0.01 &      0.01 \\

Reduction                           &    78.86 &    78.69 &     78.50 &     78.73 \\
\multicolumn{5}{l}{(TreeBagging(50) as \% of Uniform)}\\ 
\bottomrule
\end{tabular}

    \caption{This table provides the averaged per-period regret attained when data is collected using bandit algorithms (indicated in rows) with default parameters in Table \ref{tab:parameters}. Each column indicates the regularization parameter used in fitting the outcome model that underlies our simulation of participant outcomes. Lower regularization parameters correspond to higher heterogeneity in the simulated data. Averages and standard errors (in parentheses) are computed across over 1,000 simulations per cell. The last row indicates the reduction in average regret of TreeBagging(50) over Uniform.
    }
\end{table}

\section{Conclusion}
In this paper, we consider the problem of designing an adaptive experiment when the goal is to learn a personalized treatment assignment rule. Adaptive experiments risk discarding potentially successful treatment arms too early, sacrificing accuracy at the expense of minimizing cumulative regret.  We demonstrate that policy learning methods are able to find policies of higher values when data is collected through uniform assignment probabilities rather than standard contextual bandit algorithms, and we propose a simple heuristic to overcome this issue. Specifically, we impose a lower bound on assignment probabilities which decays slowly so that no arm is discarded too quickly. When conducting policy learning using the adaptively-collected data, we also compute a score that tracks how much each arm is favored by the adaptive algorithm and restrict the set of arms used to learn a policy to those that are most favored in the adaptive data collection. 

We illustrate our approach by conducting an adaptive survey experiment eliciting user preferences in charitable giving. Our setting considers data sets that cost several thousand dollars to collect using commonly accessed survey platforms, and so the size of our experiment is comparable in size to those accessible to academic researchers. The benefits to personalization are likely to be greater in our experiment than many others - it was designed with personalization in mind - so our findings may not be representative of the challenges faced by standard algorithms in more typical studies. We believe there is a great opportunity for future research to improve on the performance of adaptive experiments in settings with moderate sample sizes relative to the signal in the data.

\label{sec:conclusion}

\bibliography{references}
\bibliographystyle{apalike}

\clearpage
\appendix

\section{Survey Details}
\label{sec:contexts}

\paragraph{Contexts} The first page of our survey (after an introductory consent page) asked the following questions. Note that some of the options were binned or binarized before analysis.

\begin{itemize}
    \item \texttt{age}: How old are you?
    
    \begin{itemize}
        \item Integer
    \end{itemize}
    
    \item \texttt{male}: What is your gender?
    \begin{itemize}
        \item Alternatives: male, female, other, or prefer not to say
        \item Binarized to: male (1) or other (0)
    \end{itemize}
    
    \item \texttt{race}: Of the following options, what best describes you?
    \begin{itemize}
        \item Alternatives: White, Black or African American, American Indian or Alaska Native, Asian, Native Hawaiian or Pacific islander, Other
        \item Binarized to: white (1) or other (0)
    \end{itemize}
    
    \item \texttt{married} What option best describes your marital status? 
    \begin{itemize}
        \item Alternatives: Single,  Married, Widowed, Divorced or Separated
        \item Binarized to: married (1) or other (0)
    \end{itemize}
    
    \item \texttt{last\_donation}: When was the last time you donated to a charity?
    \begin{itemize}
        \item Alternatives: Within this month, Within this year, More than a year ago, Never
        \item Mapped to: 1-4
    \end{itemize}
    
    \item \texttt{political\_leaning}: Which political party do you identify yourself with?
    \begin{itemize}
        \item Alternatives: Strong Democrat, Moderate Democrat, Leaning Democrat, Independent/None, Leaning Republican, Moderate Republican, Strong Republican
        \item Mapped to: 1-7
    \end{itemize}
    
    \item \texttt{religious}: How religious would you consider yourself?
    \begin{itemize}
        \item Alternatives: Very religious, Moderately religious, Not religious
        \item Binned to: very/moderately (1) or not (0)
    \end{itemize}

    \item \texttt{rural}: How would you describe where you reside?
    \begin{itemize}
        \item Alternatives: Rural, Suburban, Urban
        \item Binned to: rural/suburban (1), urban (0)
    \end{itemize}
    
    \item \texttt{views\_*}: We'd like to ask some questions about your views on different policy issues. Please select the choice that best corresponds to your position.
    \begin{itemize}
        \item Questions:
        \begin{itemize}
            \item \texttt{views\_immigration}: \textit{The US gov't needs to get tougher on immigration}
            \item \texttt{views\_global\_warming}: \textit{The US gov't should do more to prevent global warming}
            \item \texttt{views\_right\_bear\_arms}: \textit{The right to bear arms should be limited}
            \item \texttt{views\_abortion}: \textit{Abortion should be banned or aggressively restricted}
        \end{itemize}
        \item Alternatives: Strongly disagree, Somewhat disagree, Neither agree nor disagree, Somewhat agree, Strongly agree
        \item Mapped to: 1-5
    \end{itemize}

    \item \texttt{news\_*}: How often do you spend time reading or watching the following news sources?
    \begin{itemize}
        \item Venues: Fox News, CNN, New York Times, Washington Post, Wall Street Journal
        \item Alternatives: Daily, Several times a week, Once a week, Several times a month, Several times a year, Once a year or less
        \item Mapped to: 1-6
    \end{itemize}
    
    \item \texttt{social\_media}: How often do you spend time on social media (Facebook, Instagram, Twitter, Reddit, etc)?
    \begin{itemize}
        \item Alternatives: Daily, Several times a week, Once a week, Several times a month, Several times year, Once a year or less
        \item Mapped to: 1-6
    \end{itemize}
\end{itemize}

\begin{figure}[H]
    \centering
    \includegraphics[width=.5\linewidth]{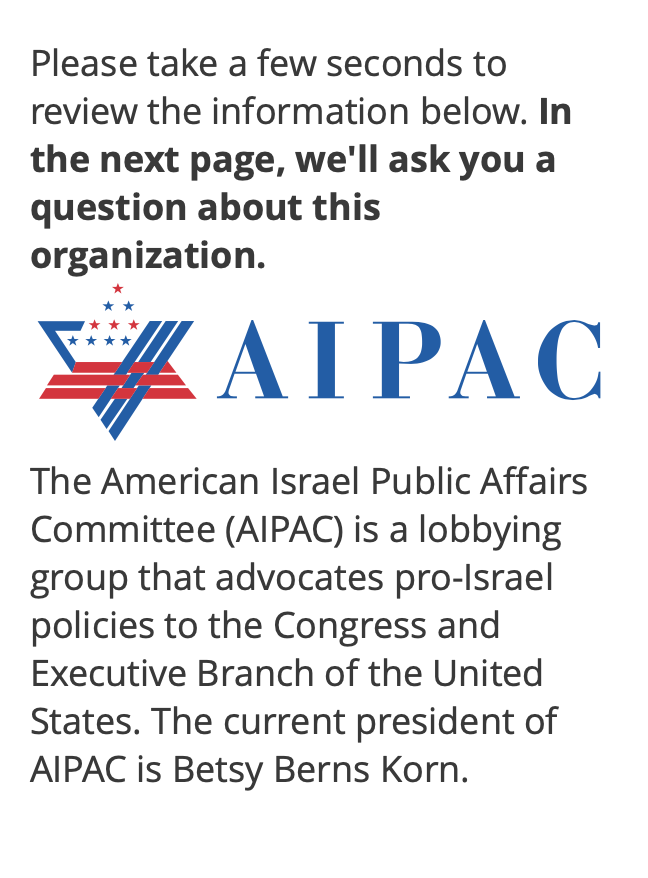}
    \caption{Screenshot of the treatment page for the \texttt{AIPAC} treatment.}
    \label{fig:treatment_screenshot}
\end{figure}

On page 2, following the algorithm described in Section \ref{sec:design}, we drew an organization that was displayed to the participant. See Figure \ref{fig:treatment_screenshot} for an example. The set of all charities is in Table \ref{tab:charities}.

\begin{table}[H]
    \centering
    \begin{tabular}{l|l} \toprule
        Charity & Alias \\ \midrule
        American Israel Public Affairs Committee &  \texttt{aipac} \\
        Black Lives Matter & \texttt{blm} \\
        Chan Zuckerberg Initiative & \texttt{zuckerberg} \\
        Clinton Foundation & \texttt{clinton} \\
        Greenpeace &  \texttt{green} \\
        National Rifle Association & \texttt{nra} \\
        People for the Ethical Treatment of Animals & \texttt{peta} \\
        Planned Parenthood & \texttt{planned} \\
        Salvation Army & \texttt{salvation} \\ \bottomrule
    \end{tabular}
    \caption{Charities and their aliases}
    \label{tab:charities}
\end{table}

\begin{figure}
    \centering
    \includegraphics[width=.8\textwidth]{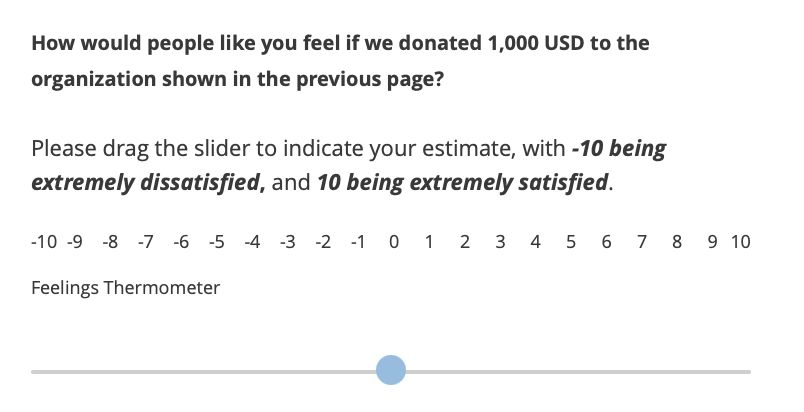}
    \caption{Screenshot of outcome question}
    \label{fig:outcome_screenshot}
\end{figure}

The third and final page asks participants two questions. First, to make sure that participants were paying attention to the treatment, we asked them to select out of a list which charity they saw in the previous page. Participants who chose a different charity than the one shown to them were dropped from the experiment. A screenshot of the final question is shown in Figure \ref{fig:outcome_screenshot}.

\section{Pilot Experiment Details} \label{sec:pilot}

\begin{table}[H]
    \centering
    \begin{tabular}{lrr}
\toprule
{} &   Mean &  Std.Error \\
Arm                        &        &            \\
\midrule
AIPAC                      & -0.159 &      0.376 \\
BLM                        &  2.027 &      0.439 \\
Clinton Foundation         & -0.285 &      0.428 \\
Greenpeace                 &  4.985 &      0.334 \\
NRA                        & -2.627 &      0.453 \\
PETA                       &  1.285 &      0.405 \\
Planned Parenthood         &  4.199 &      0.419 \\
Salvation Army             &  4.917 &      0.373 \\
Chan-Zuckerberg Initiative &  2.829 &      0.354 \\
\bottomrule
\end{tabular}

    \caption{Average response (from -10 to 10) for each charity, across all participants in Pilot 1. The means and standard errors are calculated using the full sample (n=2,463).}
    \label{tab:pilot_charity_overall}
\end{table}

\begin{table}[H]
    \hspace*{-1cm}
    \centering
    \begin{tabular}{lrrrr}
\toprule
{} & \multicolumn{2}{c}{Conservative} & \multicolumn{2}{c}{Liberal} \\
{} &         Mean & Std.Error &    Mean & Std.Error \\
Arm                        &              &           &         &           \\
\midrule
AIPAC                      &        0.893 &     0.504 &  -1.559 &     0.543 \\
BLM                        &       -1.054 &     0.625 &   5.232 &     0.470 \\
Clinton Foundation         &       -2.571 &     0.551 &   2.737 &     0.569 \\
Greenpeace                 &        3.572 &     0.521 &   6.586 &     0.348 \\
NRA                        &        0.678 &     0.590 &  -6.415 &     0.533 \\
PETA                       &        0.660 &     0.546 &   2.064 &     0.598 \\
Planned Parenthood         &        1.706 &     0.620 &   7.282 &     0.368 \\
Salvation Army             &        5.745 &     0.485 &   4.066 &     0.562 \\
Chan-Zuckerberg Initiative &        2.135 &     0.496 &   3.696 &     0.492 \\
\bottomrule
\end{tabular}

    \caption{This table shows the average response (on a scale from -10 to 10) for participants in each group in Pilot 1 (``conservatives'' include Republican- and independent-leaning participants; ``liberals'' include Democrat-leaning participants). The means and standard errors are calculated using the full sample (n=2,463).}
    \label{tab:pilot_charity_per_leaning}
\end{table}

\begin{figure}[htbp]
     \centering
     \begin{subfigure}[b]{.2\textwidth}
         \centering
         \includegraphics[width=.8\textwidth]{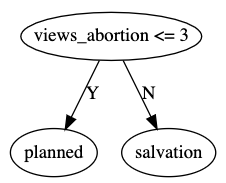}
         \caption{Maximal depth 1}
         \label{fig:pilot_tree1}
     \end{subfigure}
     \hfill
     \begin{subfigure}[b]{.35\textwidth}
         \centering
         \includegraphics[width=\textwidth]{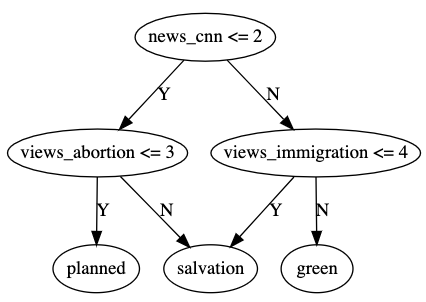}
         \caption{Maximal depth 2}
         \label{fig:pilot_tree2}
     \end{subfigure}
     \hfill
     \begin{subfigure}[b]{0.35\textwidth}
         \centering
         \includegraphics[width=\textwidth]{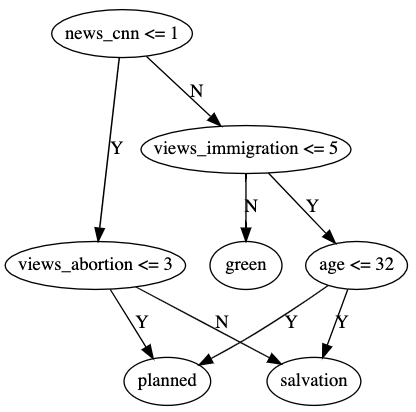}
         \caption{Maximal depth 3}
         \label{fig:pilot_tree3}
     \end{subfigure}
     \caption{Tree policies learned on the first half of pilot data in Pilot 1. The variable \texttt{views\_abortion} is a number, 1-5, ranking how much the participants think abortion should be more strictly regulated (lower numbers indicate stronger disagreement); the variable \texttt{news\_cnn} is a number, 1-6, ranking how often the participants watch CNN (3 is weekly, lower numbers indicate more frequently).}
     \label{fig:pilot_learned_policies}
\end{figure}

We use half of the pilot data to learn a non-contextual policy and tree policies of maximal depths $d=1$ to $d=3$ (a tree of depth $d$ has at most $2^{d}$ terminal nodes). Denoting by $\Pi(d)$ the set of available policies consisting of trees with maximal depth $d$, we solve the empirical problem 
\begin{equation}
  \hpi(d) = \arg\max_{\pi \in \Pi(d)} \frac{1}{n/2} \sum_{t=1}^{n/2} \widehat{Y}_t(X_t, \pi(X_t)),
\end{equation}
where $n$ is the number of observations in the pilot, and $\widehat{Y}(x, w)$ is an \emph{unbiased score} --- a transformed outcome satisfying $\EE[\widehat{Y}_t(X_t, w)|X_t] = \EE[Y_t(w)|X_t]$. The unbiased score is obtained via the AIPW technique described in Appendix \ref{sec:aipw}. The tree fitting used \texttt{R} package \texttt{policytree} \citep{sverdrup2020policytree}, and in constructing unbiased scores we used the \texttt{R} package \texttt{grf}. The learned tree policies are shown in Figure \ref{fig:pilot_learned_policies}. The non-contextual policy always assigns \texttt{Greenpeace}, which was the charity satisfying $\hpi(0) := \arg\max_{w} 1/(n/2) \sum_{t=1}^{n/2} \widehat{Y}_t(X_t, w)$.

These policies were evaluated on the remaining portion of the pilot data. Table \ref{tab:pilot_learned_policies} displays the value of each policy and the difference in value between the contextual and non-contextual policies. The results again suggest benefits from personalization, as the contextual policies attain a higher value.

\begin{table}[H]
    \centering
    \begin{tabular}{lccccc}
\toprule
{} &  Est. Value & Std. Error & Est. Diff & Std. Error & p-value \\
Policy                &             &            &           &            &         \\
\midrule
Best fixed policy (Greenpeace)  &       5.185 &      0.362 &           &            &         \\
Learned contextual policy (depth=1) &       6.889 &      0.356 &    1.704 &      0.508 &   0.001 \\
Learned contextual policy (depth=2) &       6.340 &      0.385 &    1.155 &      0.529 &   0.029 \\
Learned contextual policy (depth=3) &       6.282 &      0.409 &    1.098 &      0.546 &   0.045 \\
\bottomrule
\end{tabular}

    \caption{Value of learned policies and improvement by personalization in Pilot 1. Policies are learned on the first half of pilot data and evaluated in the remaining half via AIPW scores. Columns 1-2 show policy values, and columns 3-5 refer to the difference between contextual policies and non-contextual policies.} 
    \label{tab:pilot_learned_policies}
\end{table}

\begin{table}[H]
    \centering
    \begin{tabular}{lrr}
\toprule
{} &  Mean &  Std.Error \\
Arm                        &       &            \\
\midrule
AIPAC                      & 1.823 &      0.514 \\
BLM                        & 2.732 &      0.580 \\
Clinton Foundation         & 0.688 &      0.672 \\
Greenpeace                 & 4.153 &      0.496 \\
NRA                        & 4.809 &      0.354 \\
PETA                       & 3.267 &      0.608 \\
Planned Parenthood         & 3.628 &      0.649 \\
Salvation Army             & 6.422 &      0.110 \\
Chan-Zuckerberg Initiative & 2.043 &      0.641 \\
\bottomrule
\end{tabular}

    \caption{Average response (from -10 to 10) for each charity, across all participants in Pilot 2. The means and standard errors are calculated using the full sample (n=3,064).}
    \label{tab:pilot_charity_overall_lucid}
\end{table}

\begin{table}[H]
    \hspace*{-1cm}
    \centering
    \begin{tabular}{lrrrr}
\toprule
{} & \multicolumn{2}{l}{Conservative} & \multicolumn{2}{l}{Liberal} \\
{} &         Mean & Std.Error &    Mean & Std.Error \\
Arm                        &              &           &         &           \\
\midrule
AIPAC                      &        2.575 &     0.690 &   1.015 &     0.758 \\
BLM                        &       -0.871 &     0.831 &   6.608 &     0.534 \\
Clinton Foundation         &       -2.820 &     0.866 &   5.146 &     0.620 \\
Greenpeace                 &        3.174 &     0.662 &   6.022 &     0.612 \\
NRA                        &        6.641 &     0.306 &  -0.976 &     0.837 \\
PETA                       &        2.481 &     0.815 &   4.462 &     0.886 \\
Planned Parenthood         &        0.817 &     0.992 &   6.393 &     0.681 \\
Salvation Army             &        6.750 &     0.134 &   5.992 &     0.183 \\
Chan-Zuckerberg Initiative &        0.673 &     0.950 &   3.825 &     0.727 \\
\bottomrule
\end{tabular}

    \caption{This table shows average response (on a scale from -10 to 10) for participants in each group in Pilot 2 (``conservatives'' include Republican- and independent-leaning participants; ``liberals'' includes Democrat-leaning participants). The means and standard errors are calculated using the full sample (n=3,064).}
    \label{tab:pilot_charity_per_leaning_lucid}
\end{table}

The Pilot 2 is an adaptive experiment. In the learning phase, we collected 500 observations using TreeBagging explained in Section \ref{sec:design} and Appendix \ref{sec:algorithms}. As for other default parameter values, the decay rate on assignment probabilities lower bound is 1/8, and the number of arms selected via frequency score is 4.
\begin{table}[H]
    \centering
    \begin{tabular}{lccccc}
\toprule
{} &  Est. Value & Std. Error & Est. Diff & Std. Error & p-value \\
Policy                   &             &            &           &            &         \\
\midrule
Best fixed policy (Salvation Army) &       6.286 &      0.237 &           &            &         \\
Learned contextual policy (depth=1)    &       6.322 &      0.457 &    0.037 &      0.515 &   0.943 \\
Learned contextual policy (depth=2)   &       6.808 &      0.552 &    0.523 &      0.601 &   0.385 \\
Learned contextual policy (depth=3)    &       5.885 &      0.634 &     -0.400 &      0.677 &   0.554 \\
\bottomrule
\end{tabular}

    \caption{Value of learned policies and improvement by personalization. Policies are learned on the learning data in Pilot 2 and evaluated using the evaluation data via AIPW scores. Columns 1-2 show policy values, and columns 3-5 refer to the difference between contextual policies and non-contextual policy.} 
    \label{tab:pilot_learned_policies_lucid}
\end{table}

\begin{figure}[H]
    \centering
    \includegraphics[width=.5\textwidth]{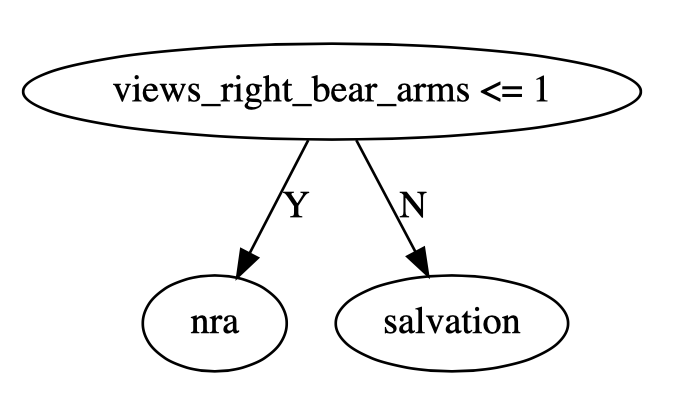}
    \caption{Tree policy learned on the learning data in Pilot 2. Maximal depth 1.}
    \label{fig:pilot_tree1_lucid}
\end{figure}

\begin{figure}[H]
    \centering
    \includegraphics[width=0.8\textwidth]{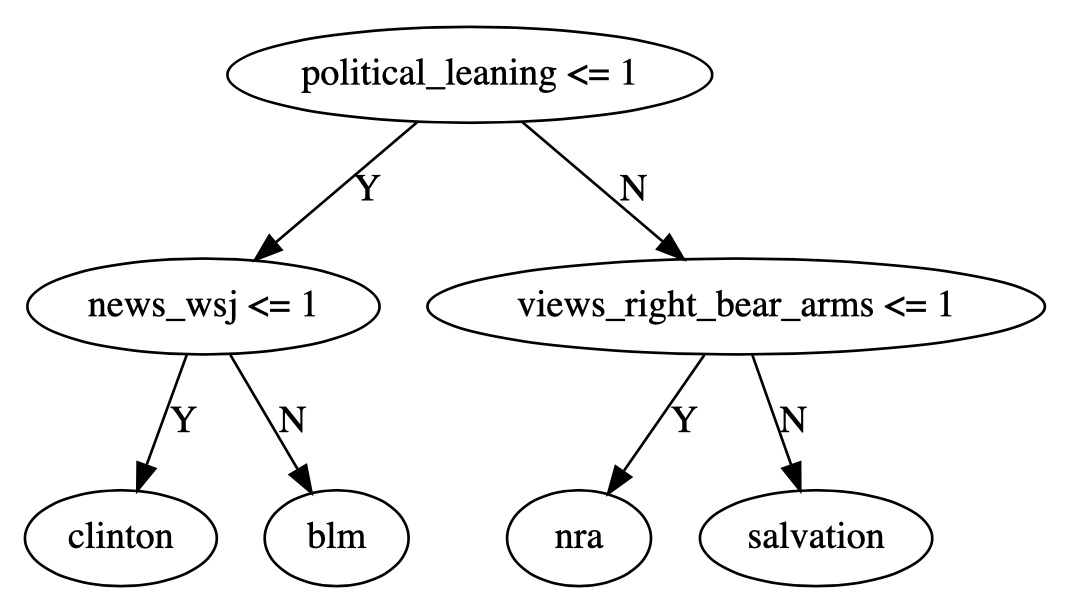}
    \caption{Tree policy learned on the learning data in Pilot 2. Maximal depth 2.}
    \label{fig:pilot_tree2_lucid}
\end{figure}

\begin{figure}[H]
    \centering
    \includegraphics[width=1\textwidth]{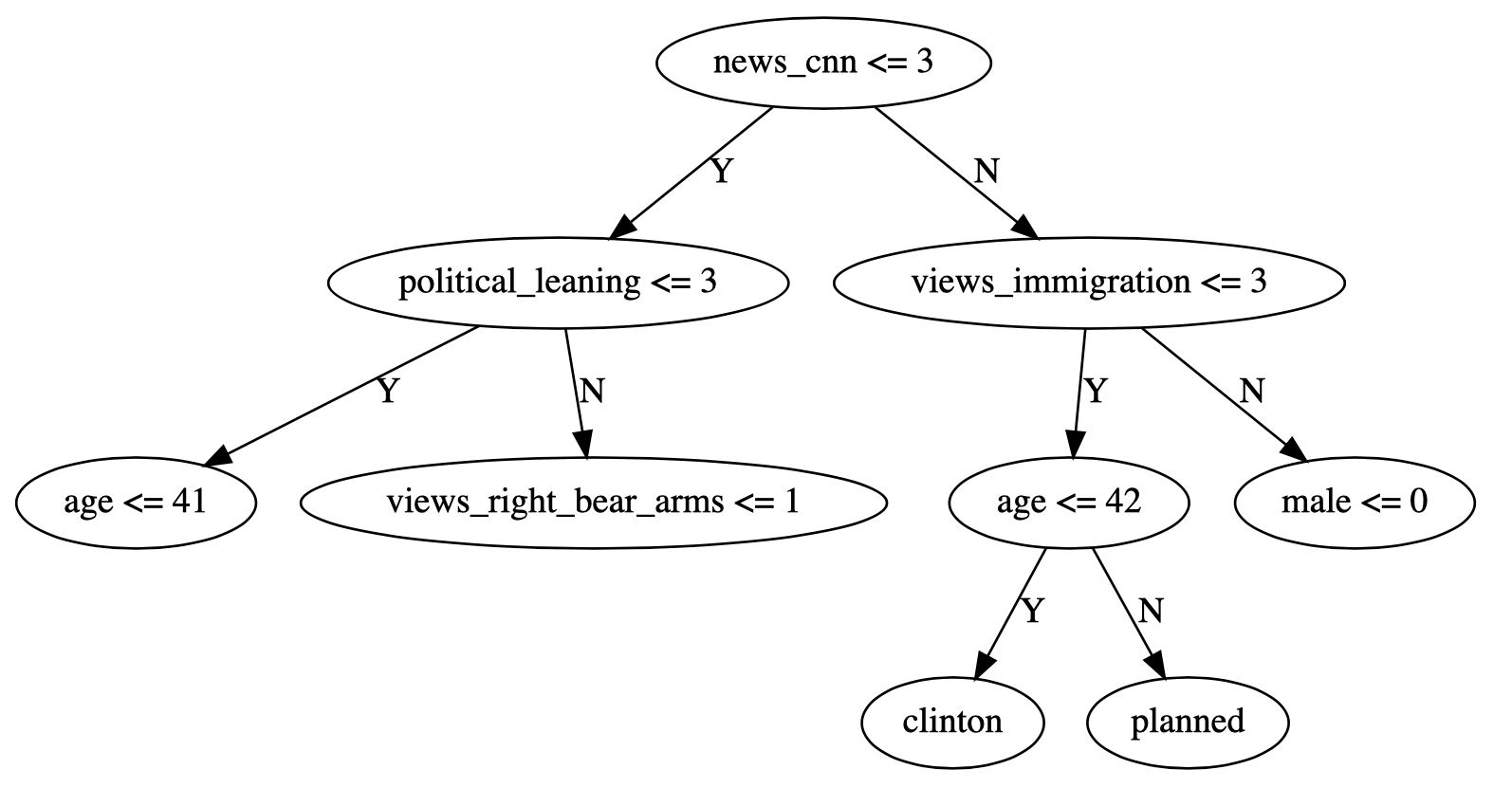}
    \caption{Tree policy learned on the learning data in Pilot 2. Maximal depth 3.}
    \label{fig:pilot_tree3_lucid}
\end{figure}

\section{Data Collection Algorithms}
\label{sec:algorithms}

\paragraph{Bootstrap Linear Thompson Sampling}

At the beginning of an adaptive batch, having collected $n$ data points. 
\begin{enumerate}
    \item Draw $M$ samples of size $n$ with replacement from current data, each denoted by $\{(X_t^{(m)}, W_t^{(m)}, Y_t^{(m)})\}$.
    
    \item Using each of the $M$ samples, for each arm $w$, estimate a linear regression model with $L_1$ penalization using only data from arm $w$,
    \begin{equation}
        \hat{\beta}_{w,m} = 
        \arg\min_{\beta} \sum_{\substack{t \leq n \\ W_t = w}}
        \left( Y_t^{(m)} - X_t^{(m)\prime}\beta \right)^2
        + \lambda \sum_{j=1}^{p} |\beta_{j}|,
    \end{equation}
    where the tuning parameter $\lambda$ can be chosen by cross-validation on the original data set.
\end{enumerate}  

At prediction time, given a next context $x$,
\begin{enumerate}
    \item Compute the mean and variance of expected reward estimates across the estimated models,
    \begin{equation}
        \hat{\mu}(x, w) := \frac{1}{M} \sum_{m=1}^M x'\hat{\beta}_{w,m} 
        \qquad
        \hat{\sigma}^2(x, w) := \frac{1}{M} \sum_{m=1}^M \left( x'\hat{\beta}_{w,m} -  \hat{\mu}(x, w) \right)^2.
    \end{equation}
    
    \item For each arm $w$, draw $N$ samples from the following normal distribution,
    \begin{equation}
        \widehat{Y}_{i,w} \sim \mathcal{N}\left(\hat{\mu}(x, w),  \hat{\sigma}^2(x, w)\right) \qquad \text{for } i = 1, ..., N.
    \end{equation}
    
    \item Tally up how often arm $w$'s draw was largest across $N$ samples,
    \begin{equation}
        \tilde{e}_t(x, w) := \frac{1}{N} \sum_{i=1}^N \mathbb{I}\{ w = \arg\max_{\tilde{w}}\widehat{Y}_{i,\tilde{w}} \}.
    \end{equation}
\end{enumerate}

\paragraph{TreeBagging} At the beginning of an adaptive batch, having collected $n$ data points. 

\begin{enumerate}
    \item Compute (possibly misspecified) estimates of the outcome model, that is, the conditional mean of rewards given contexts  $\EE[Y_t|X_t=x, W_t=w]$, denoted by $\hat{\mu}_t(x, w)$. To ensure unbiased estimates we may only use past data. In our experiment, we divide the collected data into subsets $S_1, S_2, ...$, and use observations in subsets $S_1, ..., S_{m-1}$ to produce outcome model estimates for $S_m$. Estimates for the $S_1$ subset are defaulted to zero. In our experiment, we compute estimates via the \texttt{multi\_arm\_causal\_forest} function in the R package \texttt{grf} with default tuning parameters.
    
    \item Construct AIPW scores for each arm,\footnote{In the main text, these were denoted by $\widehat{Y}_t(X_, w)$ for ease of explanation.}
    \begin{equation}
        \widehat{\Gamma}_{t}(w) :=
        \hat{\mu}_t(x, w) 
        + \frac{\mathbb{I}\{ W_t = w \}}{\tilde{e}_t(X_t, w)}
        \left( Y_t - \hat{\mu}_t(x, w)  \right) \qquad \text{ for arms }w \in [k].
    \end{equation}

    \item Draw $M$ samples of size $n$ with replacement from current data contexts and the vector of AIPW scores, denoted by $\{(X_t^{(m)}, \widehat{\Gamma}_t(\cdot)^{(m)})\}$.
    
    \item Using each of the $M$ samples, estimate policies of desired depth using the \texttt{R} package \texttt{policytree},
    \begin{equation}
        \widehat{\pi_d} := 
        \arg\max_{\pi \in \Pi(d)} \sum_{t \leq .8\tilde{T}}  \widehat{\Gamma}_{t}(\pi(X_t)),
    \end{equation}
    where $\Pi(d)$ is the set of policies of maximal depth $d$.
    
\end{enumerate}

\paragraph{Learning phase} At each batch, we compute an assignment probability function according to the following \emph{bagging} algorithm \citep[see e.g.,][]{agarwal2014taming}. At the beginning of the batch, we estimate a sequence of policies $\hat{\pi}^{(s)}(x)$ for $s \in [S]$ (we set $S=50$). Each policy in this set is fit using a sample with replacement of past data. Then, for each new value $x$ observed during the batch, we compute tentative assignment probabilities according the proportion of fitted policy in the ensemble that would assign to each arm. Letting $\tilde{e}_t(x, w)$ denote the assignment probabilities suggested by the algorithm,
\begin{equation}
    \label{eq:bagging_probs_learning}
    \tilde{e}_t(x, w) \propto | \{ \hat{\pi}^{(s)}(x) = w  \}|.
\end{equation}

\section{Policy Learning}
\label{sec:aipw}

Here we describe the steps we take to learn policies at the end of the learning phase. Let $T_{learn}$ denote the last period of the last batch of the learning phase. In our experiment, $T_{learn} = 1500$.

\begin{enumerate}
    \item Using the information from the last batch of the evaluation phase, compute the frequency score \ref{eq:freq} for each arm $w$ and select the $k$ arms with top score. In our experiment we set $k=4$. Without loss of generality here we recode the selected arms as 1-$k$.
    
    \item Drop any observations that were assigned to arms not included in the selected set above. With a little abuse of notation, for the remainder of this explanation, the time subscript $t$ from $1$ to $\tilde{T}$ should be understood as indexing only those observations.
    
    \item For every observation in the learning batch, renormalize the assignment probabilities associated with selected arms so they add up to 1,
    \begin{equation}
        \tilde{e}_t(X_t, w) :=
            \frac{e_t(X_t, w)}
                 {\sum_{w'=1}^k e_{t}(X_t, w')}.
    \end{equation}
    
    \item Compute (possibly misspecified) estimates of the outcome model, that is, the conditional mean of rewards given contexts  $\EE[Y_t|X_t=x, W_t=w]$, denoted by $\hat{\mu}_t(x, w)$. To ensure unbiased estimates we may only use past data. In our experiment, we divide the collected data into subsets $S_1, S_2, ...$ (we use subsets of size 50, though the size of the set is arbitrary), and use observations in subsets $S_1, ..., S_{m-1}$ to produce outcome model estimates for $S_m$. Estimates for the $S_1$ subset are defaulted to zero. In our experiment, we compute estimates via the \texttt{multi\_arm\_causal\_forest} function in the R package \texttt{grf} with default tuning parameters.
    
    \item Construct AIPW scores for each arm,\footnote{In the main text, these were denoted by $\widehat{Y}_t(X_, w)$ for ease of explanation.}
    \begin{equation}
        \widehat{\Gamma}_{t}(w) :=
        \hat{\mu}_t(x, w) 
        + \frac{\mathbb{I}\{ W_t = w \}}{\tilde{e}_t(X_t, w)}
        \left( Y_t - \hat{\mu}_t(x, w)  \right) \qquad \text{ for arms }w \in [k].
    \end{equation}

    \item Using the first 80\% of the data, estimate policies of depths 1-3 using the \texttt{R} package \texttt{policytree},
    \begin{equation}
        \widehat{\pi_d} := 
        \arg\max_{\pi \in \Pi(d)} \sum_{t \leq .8\tilde{T}}  \widehat{\Gamma}_{t}(\pi(X_t)),
    \end{equation}
    where $\Pi(d)$ is the set of policies of maximal depth $d$.
    
    \item Evaluate the value $V := E[Y_t(\pi_d(X_t))]$ of each learned policy in the remaining data set.
    \begin{equation}
        \widehat{V}(\hpi_d) = 
        \frac{1}{.2\tilde{T}}
       \sum_{t > .8\tilde{T}}  \widehat{\Gamma}_{t}(\pi(X_t)).
    \end{equation}
    
    \item Select the depth whose policy attains the highest value estimate,
    \begin{equation}
        d^* := \arg\max_{d} \widehat{V}(\hpi_d).
    \end{equation}
    
    \item Reestimate a policy of maximal depth $d^*$ using the entire data set,
    \begin{equation}
        \htree \equiv \widehat{\pi}_{d^*} := 
        \arg\max_{\pi \in \Pi(d)} \sum_{t \leq \tilde{T}}  \widehat{\Gamma}_{t}(\pi(X_t)),
    \end{equation}
    
    This is our output contextual policy.
    
    \item Finally, to obtain the non-contextual policy, select the arm whose average AIPW score is maximal across the entire learning data.
    \begin{equation}
         \hfixed := 
        \arg\max_{w} \sum_{t \leq \tilde{T}}  \widehat{\Gamma}_{t}(w),
    \end{equation}
    
\end{enumerate}

\end{document}